\documentclass[fleqn,usenatbib,onecolumn]{mnras}

\usepackage{newtxtext,newtxmath}
\usepackage[T1]{fontenc}

\DeclareRobustCommand{\VAN}[3]{#2}
\let\VANthebibliography\thebibliography
\def\thebibliography{\DeclareRobustCommand{\VAN}[3]{##3}\VANthebibliography}

\usepackage{graphicx}

\newcommand\vx{\mathbf{x}}
\newcommand\vy{\mathbf{y}}
\newcommand\vr{\mathbf{r}}
\newcommand\vs{\mathbf{s}}
\newcommand\vk{\mathbf{k}}
\newcommand\vkp{\mathbf{k}'}
\newcommand\vve{\mathbf{v}}
\newcommand\hx{\hat{\mathbf{x}}}
\newcommand\hy{\hat{\mathbf{y}}}
\newcommand\hz{\hat{\mathbf{z}}}
\newcommand\hr{\hat{\mathbf{r}}}
\newcommand\hs{\hat{\mathbf{s}}}
\newcommand\hk{\hat{\mathbf{k}}}
\newcommand\hn{\hat{\mathbf{n}}}
\newcommand\intx{\int \frac{d^3\vx}{V}}
\newcommand\inty{\int \frac{d^3\vy}{V}}
\newcommand\intk{\int \frac{d^3\vk}{(2\pi)^3}}
\newcommand\intkv{\int \frac{V \, d^3\vk}{(2\pi)^3}}
\newcommand\intkp{\int \frac{V \, d^3\vkp}{(2\pi)^3}}
\newcommand\intok{\int \frac{d\Omega_k}{4\pi}}
\newcommand\intor{\int \frac{d\Omega_r}{4\pi}}
\newcommand\intos{\int \frac{d\Omega_s}{4\pi}}
\newcommand\intkave{\int \frac{dk \, k^2}{2\pi^2}}
\newcommand\intkvave{\int \frac{V \, dk \, k^2}{2\pi^2}}

\title[Analytical covariance of velocity correlations]{On the correlations of galaxy peculiar velocities and their covariance}

\author[Blake \& Turner]{
Chris Blake,$^{1}$\thanks{E-mail: cblake@swin.edu.au} \&
Ryan J. Turner,$^{1}$
\\
$^{1}$Centre for Astrophysics and Supercomputing, Swinburne University of Technology, Hawthorn, VIC 3122, Australia\\
}

\date{Accepted XXX. Received YYY; in original form ZZZ}

\pubyear{2023}

\begin{document}
\label{firstpage}
\pagerange{\pageref{firstpage}--\pageref{lastpage}}
\maketitle

\begin{abstract}
Measurements of the peculiar velocities of large samples of galaxies enable new tests of the standard cosmological model, including determination of the growth rate of cosmic structure that encodes gravitational physics.  With the size of such samples now approaching hundreds of thousands of galaxies, complex statistical analysis techniques and models are required to extract cosmological information.  In this paper we summarise how correlation functions between galaxy velocities, and with the surrounding large-scale structure, may be utilised to test cosmological models.  We present new determinations of the analytical covariance between such correlation functions, which may be useful for cosmological likelihood analyses.  The statistical model we use to determine these covariances includes the sample selection functions, observational noise, curved-sky effects and redshift-space distortions.  By comparing these covariance determinations with corresponding estimates from large suites of cosmological simulations, we demonstrate that these analytical models recover the key features of the covariance between different statistics and separations, and produce similar measurements of the growth rate of structure.
\end{abstract}

\begin{keywords}
cosmology: large-scale structure of Universe -- cosmology: theory -- methods: statistical
\end{keywords}

\section{Introduction}

The relative motion of galaxies within the cosmic web of large-scale structure is a powerful probe of cosmological models.  The ``peculiar velocities'' of individual galaxies can be measured by combining their redshifts with an independent distance indicator such as the Tully-Fisher relation, Fundamental Plane, or supernova standard candle \citep{1995PhR...261..271S}.  The statistics of peculiar velocities, and their correlation with the surrounding galaxy distribution, provide important tests of the growth rate of cosmic structure and the physics of density perturbations on the largest scales \citep[e.g.,][]{2004MNRAS.347..255B, 2014MNRAS.444.3926J, 2014MNRAS.445.4267K, 2015MNRAS.450..317C, 2017JCAP...05..015H, 2017MNRAS.464.2517H, 2017MNRAS.470..445N, 2019MNRAS.486..440D, 2020MNRAS.494.3275A, 2020MNRAS.497.1275S, 2020MNRAS.498.2703B, 2023MNRAS.518.1840L, 2023MNRAS.518.2436T, 2023A&A...670L..15C}.

Direct peculiar velocity measurements are already available for tens of thousands of galaxies through samples such as the 6-degree Field Galaxy Survey \citep{2014MNRAS.445.2677S}, the Sloan Digital Sky Survey \citep{2013A&A...557A..21S, 2022MNRAS.515..953H}, and compendiums such as the Cosmic Flows catalogue \citep{2016AJ....152...50T, 2023ApJ...944...94T}.  These samples will be expanded to hundreds of thousands of galaxies through current and future observational projects such as the Dark Energy Survey Instrument \citep{2023arXiv230213760S}, the 4-metre Multi-Object Spectroscopic Telescope (4MOST) Hemisphere Survey \citep{2023Msngr.190...46T}, the Vera Rubin Observatory \citep{2017ApJ...847..128H} and the Australian Square Kilometre Array Pathfinder WALLABY survey \citep{2023MNRAS.519.4589C}.  This abundance of new data allows application of the same ``statistical machinery'' to peculiar velocity correlations as has been developed in recent years for the analysis of large galaxy redshift surveys, in which summary 2-point statistics are estimated and compared to theoretical models using large covariance matrices.

In this paper we review the suite of 2-point correlation functions that are available for quantifying peculiar velocity statistics in configuration space.  A principal challenge for statistical analysis of 2-point functions, on which we focus in this paper, is determining the covariance matrix describing the inter-correlations between the different summary statistics and separations, which is used in cosmological Bayesian likelihood analysis.  Different methods are available for evaluating the covariance matrix of such statistics.  A straight-forward approach is to generate a large ensemble of simulated datasets and use the fluctuations over these simulations to estimate the resulting covariance.  However, this approach may be limited by the extent to which the simulations accurately represent the dataset, and the large number of simulations that might be required to minimise noise in the covariance matrix as the size of data vectors grow \citep{2007A&A...464..399H}.  A second potential approach is to estimate the covariance internally from the dataset, using jack-knife or bootstrap techniques \citep[e.g.,][]{2009MNRAS.396...19N, 2020MNRAS.491.3290P, 2021MNRAS.505.5833F, 2022MNRAS.514.1289M}.  Such estimates benefit from not requiring any additional inputs, but may be limited by the capacity to create ``equivalent'' pseudo-independent sub-samples with which to estimate global statistics.  A third approach, which we test in this study, is to calculate covariance matrices analytically by propagating errors in the context of a statistical model of the dataset.  This approach may be limited by the accuracy of this statistical model, which may not necessarily reflect the detailed properties of the dataset.  However, analytical covariances are noise-free, allow likelihood analyses to proceed where large numbers of accurate mocks are unavailable, and have proved useful for combined-probe cosmological studies where the size of the data vectors are large, such as joint analyses of weak lensing and large-scale structure statistics \citep[e.g.,][]{2017MNRAS.470.2100K, 2021MNRAS.508.3125F, 2021A&A...646A.129J}.  Analytical covariances, and hybrid approaches where these covariances are calibrated using a small number of mock catalogues or jack-knife samples, have also been widely explored in large-scale structure analyses \citep[e.g.,][]{1994ApJ...426...23F, 2016MNRAS.457.1577G, 2016MNRAS.462.2681O, 2018MNRAS.479.5168B, 2019JCAP...01..016L, 2019MNRAS.487.2701O, 2020PhRvD.102l3521W, 2020MNRAS.491.3290P, 2022PhRvD.106d3515H}.

The goal of this paper is to perform new calculations of the analytical covariance between correlation functions relevant to studies of both simulated and observational peculiar velocity datasets, and to assess the accuracy of these covariances through direct comparison with an ensemble of mock catalogues.  In Sec.\ref{sec:corrstat} we review the various correlation functions which may be used in peculiar velocity studies.  In Sec.\ref{sec:anacov} we introduce the statistical model we adopt for analysing fluctuations in density and velocity samples, and outline the calculation of the analytical covariance by applying this model to correlation function estimators.  In Sec.\ref{sec:mocks} we compare the analytical covariance to a matched mock covariance, and study their relative performance in fits for the growth rate of structure.  We conclude our study in Sec.\ref{sec:conc}.

\section{Velocity correlation statistics}
\label{sec:corrstat}

In this section we introduce the different types of velocity correlation function we will study in this paper, and their relation to the underlying matter power spectrum and growth rate of structure.  We will consider correlations involving both 3D velocity vectors (which may be measured from simulations of cosmic structure and used to test theoretical models), and line-of-sight velocities (which may be measured from observational data using redshift-independent distances).

\subsection{3D density and velocity correlations}
\label{sec:corr3d}

The rate of change of a matter density perturbation $\delta_m$ with time may be related to the surrounding velocity field using the continuity equation, in terms of the growth rate of cosmic structure $f = d(\ln{\delta_m})/d(\ln{a})$, where $a$ is the cosmic scale factor.  The components of the 3D velocity field, $v_i(\vx)$, at vector position $\vx$ relative to an origin, are given in ``linear theory'' by,
\begin{equation}
    \tilde{v}_i(\vk) = -\frac{i \, k_i \, a \, H \, f}{k^2} \tilde{\delta}_m(\vk) ,
\label{eq:continuity}
\end{equation}
where $\tilde{v}_i(\vk)$ and $\tilde{\delta}_m(\vk)$ are the Fourier amplitudes of these fields in terms of wavevector $\vk$, and $H = \dot{a}/a$ is the Hubble parameter.  Eq.\ref{eq:continuity} assumes linear perturbation theory and irrotational velocity fields (for a derivation, see for example \cite{2017MNRAS.471..839A}, Appendix C).  The Fourier amplitudes of matter fluctuations are related to the underlying matter power spectrum, $P_m(k) = \langle |\tilde{\delta}_m(\vk)|^2 \rangle \, V$, which may be predicted by cosmological theories, where $V$ is the volume of the normalising Fourier cuboid and the angled brackets $\langle ... \rangle$ indicate an average over many statistical realisations.  We can trace the matter overdensity field through the locations of discrete galaxies. In a linear bias model, the galaxy overdensity at a location is $\delta_g(\vx) = b \, \delta_m(\vx)$, where $b$ is the galaxy bias factor. 

Information about the matter power spectrum and growth rate can be gleaned by measuring the auto-correlation and cross-correlation functions of the galaxy overdensity and velocity fields.  We define the galaxy auto-correlation function, $\xi_{gg}(\vr)$, as a function of vector separation $\vr$, in terms of the galaxy overdensity field as,
\begin{equation}
    \xi_{gg}(\vr) = \langle \delta_g(\vx) \, \delta_g(\vx+\vr) \rangle = \intk \, P_{gg}(k) \, e^{-i \vk \cdot \vr} ,
\label{eq:ggcorr}
\end{equation}
where its relation to the galaxy power spectrum, $P_{gg}(k) = b^2 \, P_m(k)$, follows by expressing $\delta_g(\vx)$ in terms of its Fourier components.  Neglecting redshift-space distortions, $P_{gg}(k)$ does not depend on the direction of $\vk$, through isotropy.  Given the vector nature of the velocity field, we define the velocity correlation tensor $\psi_{ij}(\vr)$ as,
\begin{equation}
    \psi_{ij}(\vr) = \langle v_i(\vx) \, v_j(\vx+\vr) \rangle = \intk \, \frac{k_i \, k_j}{k^2} \, P_{vv}(k) \, e^{-i \vk \cdot \vr} ,
\label{eq:velcorrtens}
\end{equation}
where the second equality in terms of the velocity power spectrum, $P_{vv}(k) = \frac{a^2 H^2 f^2}{k^2} \, P_m(k)$ follows by substituting in Eq.\ref{eq:continuity}.  We also define the cross-correlation between the galaxy overdensity and each velocity component, which we write as $\zeta_i(\vr)$,
\begin{equation}
    \zeta_i(\vr) = \langle \delta_g(\vx) \, v_i(\vx+\vr) \rangle = \intk \, \frac{k_i}{k} \, P_{gv}(k) \, e^{-i \vk \cdot \vr} ,
\end{equation}
where we define the galaxy-velocity cross-power spectrum, $P_{gv}(k) = \frac{i b a H f}{k} \, P_m(k)$.

We note an important practical detail that measurements of these correlation functions often trace velocities using galaxies, which are preferentially located in overdense regions, hence tracing the ``momentum'' or mass-weighted correlation function weighted by $1+\delta_g(\vx)$ \citep{2000MNRAS.319..573P, 2019MNRAS.487.5209H}.  This effect introduces higher-order corrections which we neglect in our current study, in which we choose to focus on large scales.

\subsection{Forming isotropic velocity correlations}
\label{sec:isotropic}

In order to compare correlation function measurements with theoretical predictions, it is convenient to use summary statistics which are ``isotropic'' -- that is, which depend only on the magnitude of separation $r = |\vr|$, not direction $\hr$.  In that case, we may conveniently angle-average measurements over different directions of the separation vector.  Neglecting redshift-space distortions, which we will consider in Sec.\ref{sec:rsd}, the galaxy auto-correlation is an isotropic function of $\vr$, which we can see by evaluating Eq.\ref{eq:ggcorr} as,
\begin{equation}
    \xi_{gg}(r) = \intkave \, P_{gg}(k) \intok \, e^{-i \vk \cdot \vr} = \intkave \, P_{gg}(k) \, j_0(kr) ,
\label{eq:ggcorr2}
\end{equation}
where we have used the relation for the spherical Bessel function $j_0$,
\begin{equation}
j_0(kr) = \intor \, e^{i \vk \cdot \vr}.
\end{equation}
However, $\psi_{ij}(\vr)$ and $\zeta_i(\vr)$ are not isotropic functions, and are therefore not useful as a comparison point between measurements and theory.

The velocity correlation tensor $\psi_{ij}(\vr)$ can be expressed in terms of two isotropic functions $\psi_\parallel(r)$ and $\psi_\perp(r)$ as \citep{1988ApJ...332L...7G},
\begin{equation}
    \psi_{ij}(\vr) = \psi_\perp(r) \, \delta^K_{ij} + \left[ \psi_\parallel(r) - \psi_\perp(r) \right] \, \left( \frac{r_i \, r_j}{r^2} \right) ,
\label{eq:velcorrtens2}
\end{equation}
where $\delta^K_{ij}$ is the Kronecker delta function, and the functions $\psi_\parallel(r)$ and $\psi_\perp(r)$ are given by,
\begin{equation}
  \psi_\parallel(r) = \intkave P_{vv}(k) \left[ j_0(kr) - \frac{2 j_1(kr)}{kr} \right] ,
\end{equation}
and,
\begin{equation}
  \psi_\perp(r) = \intkave P_{vv}(k) \, \frac{j_1(kr)}{kr} .
\end{equation}
We have not found a convenient derivation of Eq.\ref{eq:velcorrtens2} in the literature, so we provide one in Appendix \ref{sec:psiderivation}.  

Using these results, we can readily derive two isotropic summary statistics from the 3D velocity field \citep{1988ApJ...332L...7G}.  First, we may define the correlation function between the inward velocities between two points, $v_r(\vx) = \vve(\vx) \cdot \hr = \sum_i v_i(\vx) \, r_i/r$,
\begin{equation}
  \xi_{vv}(r) = \xi_{v_r v_r}(r) = \langle v_r(\vx) \, v_r(\vx+\vr) \rangle = \sum_{ij} \psi_{ij}(\vr) \left( \frac{r_i \, r_j}{r^2} \right) = \psi_\parallel(r) ,
\label{eq:vvcorr}
\end{equation}
where the final equality is derived by substituting in Eq.\ref{eq:velcorrtens2}.  (We emphasise that throughout this paper, $v_r$ represents the approach velocity of one galaxy towards another along separation vector $\vr$, not the radial velocity with respect to an observer that we introduce below.)  Second, we can form the isotropic total velocity correlation function,
\begin{equation}
  \xi_{v_t v_t}(r) = \sum_i \langle v_i(\vx) \, v_i(\vx+\vr) \rangle = \sum_i \psi_{ii}(\vr) = 2\psi_\perp(r) + \psi_\parallel(r) .
\end{equation}
For the galaxy-velocity cross-correlation function, we can define an isotropic correlation between galaxy position and inward velocity \citep{1995ApJ...448..494F,2014JCAP...05..003O,2017MNRAS.471..839A},
\begin{equation}
    \xi_{g v}(r) = \langle v_r(\vx) \, \delta_g(\vx+\vr) \rangle = \sum_i \frac{r_i}{r} \, \zeta_i(\vr) = \intk P_{gv}(\vk) \, (\hk \cdot \hr) = \intkave P_{gv}(k) \, j_1(kr) ,
\label{eq:gvcorr}
\end{equation}
where we have used the relation for the spherical Bessel function $j_1$,
\begin{equation}
j_1(kr) = \intor \, ( \hk \cdot \hr ) \, e^{i \vk \cdot \vr} .
\end{equation}

\subsection{Line-of-sight velocity correlations}
\label{sec:lineofsight}

When analysing observational data, we measure the line-of-sight velocity $u(\vx) = \vve(\vx) \cdot \hx = \sum_i v_i(\vx) \, x_i/x$, where $\vx$ is the vector position of the velocity tracer with respect to the observer.  Since the projection of the 3D velocity to the line-of-sight velocity depends on the position relative to the observer, the theoretical expressions now depend on the distribution of the velocity tracers through space.  We represent this by writing the correlation functions as estimators over the survey volume $V$ (this was implicitly present in our definitions of the 3D correlations in Sec.\ref{sec:corr3d} and Sec.\ref{sec:isotropic}, but not salient because the statistics were independent of position $\vx$).

We first consider the line-of-sight velocity auto-correlation function, which we estimate as,
\begin{equation}
    \hat{\xi}_{uu}(\vr) = \intx \, u(\vx) \, u(\vy) = \sum_{ij} \intx \left( \frac{x_i y_j}{xy} \right) v_i(\vx) \, v_j(\vy) ,
\end{equation}
where we write $\vy = \vx + \vr$ in this section.  Substituting in Eq.\ref{eq:velcorrtens2}, the expectation value of the line-of-sight velocity correlation function is,
\begin{equation}
\begin{split}
    \langle \hat{\xi}_{uu}(\vr) \rangle &= \sum_{ij} \intx \left( \frac{x_i y_j}{xy} \right) \psi_{ij}(\vr) = \psi_\perp(r) \left[ \intx (\hx \cdot \hy) \right] + \left[ \psi_\parallel(r) - \psi_\perp(r) \right] \left[ \intx (\hr \cdot \hx) \, (\hr \cdot \hy) \right] \\
    &= \psi_\perp(r) \, \langle \cos{\theta_{12}} \rangle + \left[ \psi_\parallel(r) - \psi_\perp(r) \right] \langle \cos{\theta_1} \cos{\theta_2} \rangle ,
\end{split}
\label{eq:uucorr}
\end{equation}
where $\langle \cos{\theta_{12}} \rangle$ represents the average of the cosine of the angle subtended by the position vectors $\vx$ and $\vy$ at the observer, and $\langle \cos{\theta_1} \cos{\theta_2} \rangle$ is the average of the product of the cosines between the position and separation vectors.  (We refer to Fig.1 of \cite{2021MNRAS.502.2087T} for a useful visualisation of the geometry.)  Note that this expression encodes the exact curved-sky effects which arise from the variation of the line-of-sight direction with position.  In a flat-sky approximation, $\cos{\theta_{12}} = 1$ and $\cos{\theta_1} = \cos{\theta_2}$.

In order to gain insight into Eq.\ref{eq:uucorr} and the nature of the $\psi_\perp(r)$ and $\psi_\parallel(r)$ functions, it is useful to consider the correlation in line-of-sight velocities of a single pair of velocity tracers in the flat-sky approximation.  If the separation vector of that pair is perpendicular to the line-of-sight, then $\cos{\theta_i} = 0$ and $\langle u_1 \, u_2 \rangle = \psi_\perp(r)$.  If the separation vector is parallel to the line-of-sight, then $\cos{\theta_i} = 1$ and $\langle u_1 \, u_2 \rangle = \psi_\parallel(r)$.  This analysis shows that the radial velocity auto-correlation function is dependent on the direction of the separation vector, and therefore angle-averaging $\hat{\xi}_{uu}(\vr)$ may lose information.

In order to extract more information from the line-of-sight velocity correlations, we may utilise the $\psi_1(r)$ and $\psi_2(r)$ statistics defined by \cite{1989ApJ...344....1G}, which apply pairwise weights for each statistic depending on the angles subtended with respect to the observer and separation vectors,
\begin{equation}
   \hat{\psi}_1(r) = N_{\psi_1} \intor \intx \, u(\vx) \, u(\vy) \left( \hx \cdot \hy \right) = N_{\psi_1} \sum_{ijk} \intx \left( \frac{x_i x_k}{x^2} \right) \left( \frac{y_j y_k}{y^2} \right) v_i(\vx) \, v_j(\vy) ,
\label{eq:psi1est}
\end{equation}
and,
\begin{equation}
\hat{\psi}_2(r) = N_{\psi_2} \intor \intx \, u(\vx) \, u(\vy) \left( \hx \cdot \hr \right) \left( \hy \cdot \hr \right) = N_{\psi_2} \sum_{ijkl} \left( \frac{r_k r_l}{r^2} \right) \intx \left( \frac{x_i x_k}{x^2} \right) \left( \frac{y_j y_l}{y^2} \right) v_i(\vx) \, v_j(\vy) ,
\label{eq:psi2est}
\end{equation}
where $N_{\psi_1}$ and $N_{\psi_2}$ are normalisation constants we identify below.  Using Eq.\ref{eq:velcorrtens2}, the expectation value of Eq.\ref{eq:psi1est} is given by,
\begin{equation}
   \langle \hat{\psi}_1(r) \rangle = N_{\psi_1} \left\{ \, \psi_\perp(r) \, \langle \cos^2{\theta_{12}} \rangle + \left[ \psi_\parallel(r) - \psi_\perp(r) \right] \langle \cos{\theta_1} \cos{\theta_2} \cos{\theta_{12}} \rangle \right\} .
\end{equation}
Defining for each separation bin the normalisation, $N_{\psi_1} = \frac{1}{\langle \cos^2{\theta_{12}} \rangle}$, and the geometry factor, $A(r) = \frac{\langle \cos{\theta_1} \cos{\theta_2} \cos{\theta_{12}} \rangle}{\langle \cos^2{\theta_{12}} \rangle}$, we find,
\begin{equation}
    \langle \hat{\psi}_1(r) \rangle = A(r) \, \psi_\parallel(r) + \left[ 1-A(r) \right] \, \psi_\perp(r) .
\label{eq:psi1corr}
\end{equation}
Likewise for $\psi_2$,
\begin{equation}
   \langle \hat{\psi}_2(r) \rangle = \left\{ \, \psi_\perp(r) \, \langle \cos{\theta_{12}} \, \cos{\theta_1} \, \cos{\theta_2} \rangle + \left[ \psi_\parallel(r) - \psi_\perp(r) \right] \langle \cos^2{\theta_1} \cos^2{\theta_2} \rangle \right\} .
\end{equation}
Defining the normalisation, $N_{\psi_2} = \frac{1}{\langle \cos{\theta_{12}} \, \cos{\theta_1} \, \cos{\theta_2} \rangle}$, and the geometry factor, $B(r) = \frac{\langle \cos^2{\theta_1} \cos^2{\theta_2} \rangle}{\langle \cos{\theta_{12}} \, \cos{\theta_1} \, \cos{\theta_2} \rangle}$, we find,
\begin{equation}
    \langle \hat{\psi}_2(r) \rangle = B(r) \, \psi_\parallel(r) + \left[ 1 - B(r) \right] \, \psi_\perp(r) .
\label{eq:psi2corr}
\end{equation}
In the flat-sky approximation, $\psi_1(r)$ and $\psi_2(r)$ are linear combinations of the first two multipoles of the expansion of $\xi_{uu}(\vr)$ about the angle to the line-of-sight ($\psi_1 = \xi^0_{uu}$, $\psi_2 = \xi^0_{uu} + \frac{2}{5} \xi^2_{uu}$), which encode all the angular information.

We now consider the cross-correlation function between the line-of-sight velocity and galaxy overdensity.  Since its monopole is zero, we consider the dipole contribution, which \cite{2021MNRAS.502.2087T} defined as,
\begin{equation}
    \hat{\psi}_3(r) = N_{\psi_3} \intor \intx \, u(\vx) \, \delta_g(\vy) \, (\hx \cdot \hr) .
\label{eq:psi3est}
\end{equation}
Using Eq.\ref{eq:gvcorr}, this function has expectation value,
\begin{equation}
    \langle \hat{\psi}_3(r) \rangle = N_{\psi_3} \, \xi_{gv}(r) \, \langle \cos^2{\theta_{12}} \rangle .
\end{equation}
Hence, with the identification $N_{\psi_3} = \frac{1}{\langle \cos^2{\theta_{12}} \rangle}$, we have $\langle \psi_3(r) \rangle = \xi_{gv}(r)$.

\subsection{Redshift-space distortions}
\label{sec:rsd}

The isotropy of correlation functions is broken in observational datasets by the effects of redshift-space distortions (RSD), through which the apparent positions of galaxies, as determined by their redshifts, are shifted according to their line-of-sight velocities.  The presence of RSD requires us to move to an isotropic basis of correlation function multipoles.  RSD has no effect on the velocity auto-correlation function to first order \citep{2021JCAP...09..018D}, hence we only consider redshift-space modifications to the cross-correlation and the galaxy auto-correlation functions.  The multipole estimators are, for the galaxy auto-correlation function,
\begin{equation}
    \hat{\xi}^\ell_{gg}(r) = \left( 2\ell + 1 \right) \intor \intx \, \delta_g(\vx) \, \delta_g(\vy) \, L_\ell(\hx \cdot \hr) ,
\label{eq:xiggest}
\end{equation}
and for the cross-correlation function,
\begin{equation}
    \hat{\xi}^\ell_{gu}(r) = \left( 2\ell + 1 \right) \intor \intx \, u(\vx) \, \delta_g(\vy) \, L_\ell(\hx \cdot \hr) ,
\label{eq:xiguest}
\end{equation}
(where we can identify $\xi^1_{gu} = \psi_3$).

To develop the relations between these correlation function multipoles and the underlying power spectra, it is convenient to express the dependence of the power spectra on the angle with respect to the line-of-sight using multipoles, such that the power spectra become position-dependent \citep[e.g,][]{2019MNRAS.489..153B},
\begin{equation}
    P_{gg}(\vk, \vx) = \sum_\ell P_{gg}^\ell(k) \, L_\ell(\hk \cdot \hx) ; \hspace{1cm} P_{gv}(\vk, \vx) = \sum_\ell P_{gv}^\ell(k) \, L_\ell(\hk \cdot \hx) .
\end{equation}
The expectation values of these estimators are given by, for the galaxy auto-correlation function multipoles \citep[e.g.,][]{2017MNRAS.469.1369S},
\begin{equation}
    \langle \hat{\xi}^\ell_{gg}(r) \rangle = i^\ell \intkave j_\ell(kr) \, P_{gg}^\ell(k) ,
\label{eq:ggcorrmult}
\end{equation}
and for the cross-correlation multipoles \citep{2020MNRAS.494.3275A,2023MNRAS.518.1840L,2023MNRAS.518.2436T},
\begin{equation}
    \langle \hat{\xi}^\ell_{gu}(r) \rangle = i^{\ell+1} \intkave j_{\ell}(kr) \sum_{\ell'} P_{gv}^{\ell'}(k) \sum_{\ell''=|\ell'-1|}^{\ell'+1} A_{\ell',1}^{\ell''} \, \delta_{\ell \, \ell''} ,
\label{eq:gucorrmult}
\end{equation}
with coefficients,
\begin{equation}
  A_{\ell,\ell'}^{\ell''} = \left(
  \begin{array}{ccc}
    \ell & \ell' & \ell'' \\
    0 & 0 & 0
  \end{array}
  \right)^2 \, (2\ell'' + 1) ,
\label{eq:coeff}
\end{equation}
where the matrix in Eq.\ref{eq:coeff} is a Wigner 3j-symbol.  For reference we give the linear-theory terms entering these predictions \citep{2020MNRAS.494.3275A}:
\begin{equation}
\begin{split}
    P_{gg}^0(k) &= \left( b^2 + \frac{2}{3} bf + \frac{1}{5} f^2 \right) \, P_m(k) ; \hspace{1cm} P_{gg}^2(k) = \left( \frac{4}{3} bf + \frac{4}{7} f^2 \right) \, P_m(k) ; \hspace{1cm} P_{gg}^4(k) = \frac{8}{35} f^2 \, P_m(k) ; \\
    P_{gv}^0(k) &= \frac{aH}{k} \left( bf + \frac{1}{3} f^2 \right) \, P_m(k) ; \hspace{1cm} P_{gv}^2(k) = \frac{aH}{k} \left( \frac{2}{3} f^2 \right) \, P_m(k) .
\end{split}
\label{eq:pkpole}
\end{equation}
In our study we focus on the subset of correlation function multipoles which can currently be measured with high signal-to-noise: $\xi_{gg}^0$, $\xi_{gg}^2$ and $\xi_{gu}^1$.

In the following analysis, we hence consider two groups of correlation functions.  First, ``simulation'' correlations, in which we utilise 3D velocities and analyse the three correlation functions $[ \xi_{gg} , \xi_{gv} , \xi_{vv} ]$.  Second, ``observation'' correlations, in which we utilise line-of-sight velocities and redshift-space multipoles, and analyse the five correlation functions $[ \xi^0_{gg} , \xi^2_{gg} , \xi^1_{gu} , \psi_1, \psi_2 ]$.

\section{Analytical covariance calculation}
\label{sec:anacov}

In this section we model the analytical covariance between the correlation statistics described in Sec.\ref{sec:corrstat}.  In general, the covariance of two correlation functions $\xi_1(r)$ and $\xi_2(s)$ evaluated at separations $r$ and $s$ is defined by,
\begin{equation}
    {\rm Cov} \left[ \xi_1(r) , \xi_2(s) \right] = \langle \xi_1(r) \, \xi_2(s) \rangle - \langle \xi_1(r) \rangle \, \langle \xi_2(s) \rangle .
\label{eq:covana}
\end{equation}
Since individual correlation statistics are a 2-point function, Eq.\ref{eq:covana} demonstrates that covariances are a 4-point function.  However, we can use approximations to reduce this 4-point function to evaluations over 2-point functions.  We illustrate how survey selection functions and noise may be included in correlation models, and perform the covariance evaluation for an example case.  We note that a full listing of our analytical covariance results is provided in Appendix \ref{sec:allthecovs}.

\subsection{Including selection functions and measurement noise}

In order to evaluate the covariance of a correlation measurement in a realistic scenario, we need to adapt our correlation models to include the effects of the survey selection function, measurement noise, and any position-dependent optimal weights that are applied.  We model two separate discrete datasets: a ``density'' sample which is used to trace the galaxy overdensity $\delta_g$, and a ``velocity'' sample which is used to probe the 3D or radial velocities.  We assume that:
\begin{itemize}
    \item The density and velocity samples may have different galaxy number densities, which vary with location as $n_g(\vx)$ and $n_v(\vx)$.
    \item Discrete density and velocity tracers are assigned position-dependent weights $w_g(\vx)$ and $w_v(\vx)$.
    \item Measurement of the galaxy overdensity and velocity fields contains Poisson noise from discrete objects.
    \item Tracer velocities contain measurement noise, which may be applied either to all velocity components (for a simulation), or in the line-of-sight direction (for observations).
\end{itemize}
In this context, the correlation between the (weighted, noisy) galaxy overdensity field at two positions $\vx$ and $\vy$ becomes \citep{1994ApJ...426...23F},
\begin{equation}
    \langle \delta_g(\vx) \, \delta_g(\vy) \rangle = f_g(\vx) \, f_g(\vy) \, \xi(\vx-\vy) + \sigma_g^2(\vx) \, \delta_D(\vx-\vy) ,
\label{eq:ggcorrmod}
\end{equation}
where $f_g(\vx) = w_g(\vx) \, n_g(\vx) \, V$ is the selection function of the density tracers.  We model the uncertainty in the galaxy overdensity using Poisson noise, for which $\sigma_g(\vx) = w_g(\vx) \sqrt{n_g(\vx) \, V}$ \citep{1994ApJ...426...23F,2019MNRAS.489..153B}.  We note here that through the inclusion of the normalising volume $V$, $f_g(\vx)$ and $\sigma_g(\vx)$ are dimensionless functions. The equivalent expression for the correlation of two velocity components $v_i(\vx)$ and $v_j(\vy)$ is,
\begin{equation}
    \langle v_i(\vx) \, v_j(\vy) \rangle = f_v(\vx) \, f_v(\vy) \, \psi_{ij}(\vx-\vy) + \sigma_v^2(\vx) \left( \frac{x_i \, x_j}{x^2} \right) \, \delta_D(\vx-\vy) .
\label{eq:vvcorrmod}
\end{equation}
Here, $f_v(\vx) = w_v(\vx) \, n_v(\vx) \, V$ is the selection function of the velocity tracers, and $\sigma_v(\vx) = \epsilon_v(\vx) \, w_v(\vx) \, \sqrt{n_v(\vx) \, V}$ is the velocity noise at a given location as a function of the individual-tracer velocity error $\epsilon_v(\vx)$, where this noise is applied in the radial direction from an observer.  This last statement can be established by considering,
\begin{equation}
    \langle u(\vx) \, u(\vy) \rangle = \sum_{ij} \frac{x_i \, x_j}{x^2} \langle v_i(\vx) \, v_j(\vy) \rangle = \sigma_v^2(\vx) \, \delta_D(\vx-\vy) \sum_{ij} \frac{x_i^2 \, x_j^2}{x^4} = \sigma_v^2(\vx) \, \delta_D(\vx-\vy) ,
\end{equation}
where the final step uses $\sum_i \frac{x_i^2}{x^2} = 1$.  If velocity noise is instead applied independently to each velocity component (for example, the non-linear contribution to the velocities in a simulation), the final term in Eq.\ref{eq:vvcorrmod} would be $\sigma_v^2(\vx) \, \delta^K_{ij} \, \delta_D(\vx-\vy)$.  Finally, the expression for the correlation between the galaxy overdensity field $\delta_g(\vx)$ and a component of velocity $v_i(\vy)$ is,
\begin{equation}
    \langle \delta_g(\vx) \, v_i(\vy) \rangle = f_g(\vx) \, f_v(\vy) \, \zeta_i(\vx-\vy) .
\label{eq:gvcorrmod}
\end{equation}
In this case there is no noise term, because the noise contribution to $\delta_g$ and $v_i$ is uncorrelated.

To calculate the analytical covariance we will apply several approximations, stated as follows:
\begin{itemize}
    \item We will assume that the galaxy overdensity and velocity fields are Gaussian random fields.
    \item We will neglect higher-order correlations; for example, the fact that velocity tracers preferentially sample overdense locations.
    \item We will neglect the variation of the survey selection function and noise fields on the scale of the separation vector.
    \item When correlating line-of-sight velocities we will assume the ``local plane-parallel'' approximation, that the position vectors of points on the scale of the separation are parallel (where this direction varies for different pairs).
\end{itemize}
These approximations become increasingly inaccurate for small separations, or when the separations are large compared to the radial distances.  However, we will show in the remainder of this paper that these approximations produce analytical covariances with a significant range of validity.

\subsection{Estimators in terms of fields}

Applying these approximations, we restate our estimators of the correlation function statistics in terms of the noisy, weighted fields, in the form that we will use to compute the analytical covariances.  Starting with the galaxy correlation function we have,
\begin{equation}
    \hat{\xi}_{gg}(\vr) = N_{\xi_{gg}} \, \intx \, \delta_g(\vx) \, \delta_g(\vy) ,
\end{equation}
where $N_{\xi_{gg}}$ is a normalisation constant.  Using Eq.\ref{eq:ggcorrmod}, we can confirm that $\langle \hat{\xi}_{gg} \rangle = \xi_{gg}$ if $N_{\xi_{gg}} = \intx \, f_g(\vx) \, f_g(\vx+\vr) \approx \intx f_g^2(\vx)$, where we have applied the approximation that the selection function does not vary on the separation scale.  We can average the above estimator over directions by writing,
\begin{equation}
\hat{\xi}_{gg}(r) = \intor \, \hat{\xi}_{gg}(\vr) .
\end{equation}
Generalising to the galaxy correlation function multipoles,
\begin{equation}
    \hat{\xi}^\ell_{gg}(\vr) = N_{\xi_{gg}} \, \left( 2\ell + 1 \right) \intx \, \delta_g(\vx) \, \delta_g(\vy) \, L_\ell(\hx \cdot \hr) .
\end{equation}
For the inward velocity auto-correlation function we can similarly write,
\begin{equation}
    \hat{\xi}_{vv}(\vr) = N_{\xi_{vv}} \, \intx \, \left[ \vve(\vx) \cdot \hr \right] \, \left[ \vve(\vy) \cdot \hr \right] = N_{\xi_{vv}} \sum_{ij} \left( \frac{r_i \, r_j}{r^2} \right) \intx \, v_i(\vx) \, v_j(\vy) ,
\end{equation}
where using the same approximation as above, $N_{\xi_{vv}} \approx \intx f_v^2(\vx)$.  For the cross-correlation between inward velocity and galaxy position,
\begin{equation}
    \hat{\xi}_{gv}(\vr) = N_{\xi_{gv}} \, \intx \, \left[ \vve(\vx) \cdot \hr \right] \, \delta_g(\vy) ,
\end{equation}
where $N_{\xi_{gv}} = \intx f_g(\vx) \, f_v(\vx)$.  We now turn to the correlations involving line-of-sight velocities.  The estimators for the $\psi_1$, $\psi_2$ and $\psi_3$ statistics have already been provided as Eq.\ref{eq:psi1est}, Eq.\ref{eq:psi2est} and Eq.\ref{eq:psi3est} above where, for evaluating the covariance, we additionally apply the local plane-parallel approximation $\hx \approx \hy$.  For the cross-correlation multipole statistics we use,
\begin{equation}
    \hat{\xi}^\ell_{gu}(\vr) = N_{\psi_3} \, \left( 2\ell + 1 \right) \intx \, u(\vx) \, \delta_g(\vy) \, L_\ell(\hx \cdot \hr) .
\end{equation}
These estimators form the basis of our analytical covariance evaluations.

\subsection{Example evaluation}

We illustrate the calculation of the analytical covariance using the example of $\psi_1$.  First, we express the covariance in terms of moments of the fields, using the estimator defined in Eq.\ref{eq:psi1est}, noting in the following that $\vy$ is now a general position vector and does not satisfy $\vy = \vx + \vr$:
\begin{equation}
\begin{split}
    {\rm Cov}[\hat{\psi}_1(\vr), \hat{\psi}_1(\vs)] &= \langle \hat{\psi}_1(\vr) \, \hat{\psi}_1(\vs) \rangle - \langle \hat{\psi}_1(\vr) \rangle \, \langle \hat{\psi}_1(\vs) \rangle \\
    &= N_{\psi_1}^2 \sum_{ijkl} \intx \inty \frac{x_i x_j y_k y_l}{x^2 y^2} \left[ \langle v_i(\vx) \, v_j(\vx+\vr) \, v_k(\vy) \, v_l(\vy+\vs) \rangle - \langle v_i(\vx) \, v_j(\vx+\vr) \rangle \, \langle v_k(\vy) \, v_l(\vy+\vs) \rangle \right] \\
    &= N_{\psi_1}^2 \sum_{ijkl} \intx \inty \frac{x_i x_j y_k y_l}{x^2 y^2} \left[ \langle v_i(\vx) \, v_k(\vy) \rangle \, \langle v_j(\vx+\vr) \, v_l(\vy+\vs) \rangle + \langle v_i(\vx) \, v_l(\vy+\vs) \rangle \, \langle v_j(\vx+\vr) \, v_k(\vy) \rangle \right] ,
\end{split}
\end{equation}
where the last step uses Wick's theorem for a Gaussian random field.  We now substitute in the correlation model from Eq.\ref{eq:vvcorrmod}, where we will just show the calculation of the first term for clarity, noting that the second term produces an identical result where $\vr-\vs$ is replaced by $\vr+\vs$:
\begin{equation}
\begin{split}
    {\rm Cov}[\hat{\psi}_1(\vr), \hat{\psi}_1(\vs)] &= N_{\psi_1}^2 \sum_{ijkl} \intx \inty \frac{x_i x_j y_k y_l}{x^2 y^2} f_v^2(\vx) \, f_v^2(\vy) \, \psi_{ik}(\vx-\vy) \, \psi_{jl}(\vx-\vy+\vr-\vs) \\
    &+ N_{\psi_1}^2 \sum_{ijkl} \intx \, f_v^2(\vx) \, \sigma_v^2(\vx) \, \left[ \frac{x_i x_j^2 x_k x_l^2}{x^6} \psi_{ik}(\vr-\vs) + \frac{x_i^2 x_j x_k^2 x_l}{x^6} \psi_{jl}(\vr-\vs) \right] \\
    &+ N_{\psi_1}^2 \sum_{ijkl} \intx \frac{x_i^2 x_j^2 x_k^2 x_l^2}{x^8} \sigma_v^4(\vx) \, \delta_D(\vr-\vs) ,
\end{split}
\end{equation}
where we have again applied the approximation that the selection function and noise do not vary on the scale of the separations, i.e. $f_v(\vx+\vr) \approx f_v(\vx)$ and $\sigma_v(\vx+\vr) \approx \sigma_v(\vx)$.  Now, we convert the evaluation to Fourier space, by substituting in the expression for the velocity correlation tensor from Eq.\ref{eq:velcorrtens} and for the delta function, $\delta_D(\vr) = \intkv e^{-i \vk \cdot \vr}$:
\begin{equation}
\begin{split}
    {\rm Cov}[\hat{\psi}_1(\vr), \hat{\psi}_1(\vs)] &= N_{\psi_1}^2 \sum_{ijkl} \intx \inty \frac{x_i x_j y_k y_l}{x^2 y^2} f_v^2(\vx) f_v^2(\vy) \intkv \frac{k_i k_j k_k k_l}{k^4} \frac{P^2_{vv}}{V^2}(\vk) e^{-i \vk \cdot (\vr-\vs)} \intkp e^{i(\vk-\vkp) \cdot (\vx-\vy)} \\
    &+ 2 \, N_{\psi_1}^2 \sum_{ijkl} \intx \, f_v^2(\vx) \, \sigma_v^2(\vx) \, \frac{x_i x_j^2 x_k x_l^2}{x^6} \intkv \, \frac{k_i k_k}{k^2} \, \frac{P_{vv}(\vk)}{V} \, e^{-i \vk \cdot (\vr-\vs)} \\
    &+ N_{\psi_1}^2 \sum_{ijkl} \intx \frac{x_i^2 x_j^2 x_k^2 x_l^2}{x^8} \sigma_v^4(\vx) \, \intkv e^{-i \vk \cdot (\vr-\vs)} ,
\end{split}
\label{eq:psi1cov}
\end{equation}
where in the first term, we make the approximation that the power spectrum is ``slowly varying'' and can be taken outside the integral over $\vkp$, producing a delta function $\delta_D(\vx-\vy)$.  We note that in Eq.\ref{eq:psi1cov}, and generally in our covariance expressions, we group the variables in combination where volume units cancel out.  We also combine the sums in Eq.\ref{eq:psi1cov} into dot products ($\sum_i \frac{x_i^2}{x^2} = 1$ and $\sum_i \frac{x_i}{x} \frac{k_i}{k} = \hx \cdot \hk$), such that,
\begin{equation}
\begin{split}
    {\rm Cov}[\hat{\psi}_1(\vr), \hat{\psi}_1(\vs)] &= N_{\psi_1}^2 \intkv \, \frac{P_{vv}^2(\vk)}{V^2} \, e^{i\vk \cdot (\vr - \vs)} \intx \, (\hx \cdot \hk)^4 \, f_v^4(\vx) \\
    &+ 2 \, N_{\psi_1}^2 \intkv \, \frac{P_{vv}(\vk)}{V} \, e^{i\vk \cdot (\vr - \vs)} \intx \, (\hx \cdot \hk)^2 \, f_v^2(\vx) \, \sigma_v^2(\vx) + N_{\psi_1}^2 \intkv \, e^{i\vk \cdot (\vr - \vs)} \intx \, \sigma_v^4(\vx) .
\end{split}
\end{equation}
This expression can be written more compactly as,
\begin{equation}
    {\rm Cov} \left[ \hat{\psi}_1(\vr) , \hat{\psi}_1(\vs) \right] = N_{\psi_1}^2 \intkv e^{-i \vk \cdot (\vr - \vs)} \intx \, \left[ f^2_v(\vx) \, \frac{P_{vv}(\vk)}{V} \, ( \hx \cdot \hk )^2 + \sigma_v^2(\vx) \right]^2 .
\end{equation}
Finally, we angle-average the measurements,
\begin{equation}
\begin{split}
    {\rm Cov} \left[ \hat{\psi}_1(r) , \hat{\psi}_1(s) \right] &= \intor \intos \, {\rm Cov} \left[ \hat{\psi}_1(\vr) , \hat{\psi}_1(\vs) \right] \\
    &= N_{\psi_1}^2 \intkv \intor e^{-i \vk \cdot \vr} \intos \, e^{-i \vk \cdot \vs} \intx \, \left[ f^2_v(\vx) \, \frac{P_{vv}(\vk)}{V} \, ( \hx \cdot \hk )^2 + \sigma_v^2(\vx) \right]^2 \\
    &= N_{\psi_1}^2 \intkvave \, j_0(kr) \, j_0(ks) \left[ W_4(f_v^4) \, \frac{P_{vv}^2(k)}{V^2} + 2 \, W_2(f_v^2 \sigma_v^2) \, \frac{P_{vv}(k)}{V} + W_0(\sigma_v^4) \right] ,
\end{split}
\end{equation}
where we have introduced the averages of the survey selection and noise functions over the survey window,
\begin{equation}
    W_n(f) = \intx \, f(\vx) \intok ( \hx \cdot \hk )^n .
\label{eq:selave}
\end{equation}
We do not reproduce the calculation of the other covariance elements, but rather state the final results in Appendix \ref{sec:allthecovs}.

\subsection{Scaling parameters}

The accuracy of analytical covariance matrices may be improved by the inclusion of empirical scaling parameters, which can be set through comparison with a small number of mock catalogues, or through internal error determinations such as jack-knife sampling \citep[e.g.,][]{2016MNRAS.462.2681O, 2020MNRAS.491.3290P}.  These scaling parameters can improve some of the approximations used when generating analytical covariances, such as the choice of the effective galaxy bias value, or the assumption of Poisson noise for the galaxy distribution, or Gaussian noise in the velocities.  We introduced separate empirical scaling parameters for the amplitude of the galaxy and velocity sample variance ($\alpha_{P,g}$ and $\alpha_{P,v}$) and noise variance ($\alpha_{N,g}$ and $\alpha_{N,v}$) entering the analytical covariances, and set these parameters through comparison with a subset of mock catalogues.  Referring to the covariance formulae provided in Appendix \ref{sec:allthecovs}, these scaling parameters multiply terms in $(P_{gg}; P_{gv}; P_{vv})$ by $(\alpha_{P,g}; \sqrt{\alpha_{P,g} \alpha_{P,v}}; \alpha_{P,v})$, and terms in $(\sigma^2_g; \sigma^2_v)$ by $(\alpha_{N,g}; \alpha_{N,v})$.  We found that the preferred scaling parameters increased the velocity noise variance by $30\%$ from our fiducial calculation ($\alpha_{N,v} \approx 1.15$), and changed the amplitude of the other terms by less than $10\%$ ($|\alpha-1| < 0.05$).  We included these scaling parameters in the following analysis, although note that they do not significantly impact the best-fitting model parameters in any case.

\section{Tests using simulations}
\label{sec:mocks}

In this section we use N-body simulations to test the accuracy with which the methods of Sec.\ref{sec:anacov} can predict the covariance of correlation function statistics, by estimating these covariances using measurements across many realisations of a cosmological simulation as,
\begin{equation}
    \hat{\rm Cov} \left[ \xi_1(r) , \xi_2(s) \right] = \frac{N_{\rm mock}}{N_{\rm mock} - 1} \left[ \overline{\xi_1(r) \, \xi_2(s)} - \overline{\xi_1(r)} \cdot \overline{\xi_2(s)} \right] ,
\end{equation}
where the overbar indicates an average over $N_{\rm mock}$ realisations.  We will compare the values of diagonal and off-diagonal elements of the analytical and mock covariances, as well as the performance of the covariance matrices in fits for the growth rate of structure parameter.

\subsection{Mocks}

For the purposes of these tests we used the large suite of N-body simulations described by \cite{2016MNRAS.459.2118K}, which was created using the Comoving Lagrangian Acceleration (COLA) technique.  These simulations span an $L = 600 \, h^{-1} \, {\rm Mpc}$ box, in which we placed an observer at the central location, creating an initial spherical selection function within a $300 \, h^{-1} \, {\rm Mpc}$ radius.  The fiducial power spectrum of the simulations is the ``WMAP5'' cosmological model with matter density $\Omega_m = 0.273$, baryon density $\Omega_b = 0.0456$, Hubble parameter $h = 0.705$, spectral index $n = 0.961$, and power spectrum normalisation $\sigma_8 = 0.812$.  The simulation resolves cosmological halos with masses below $M = 10^{12} \, h^{-1} \, M_\odot$, and we based our analysis on these halo catalogues.

\begin{figure}
\begin{center}
  \includegraphics[width=0.6\columnwidth]{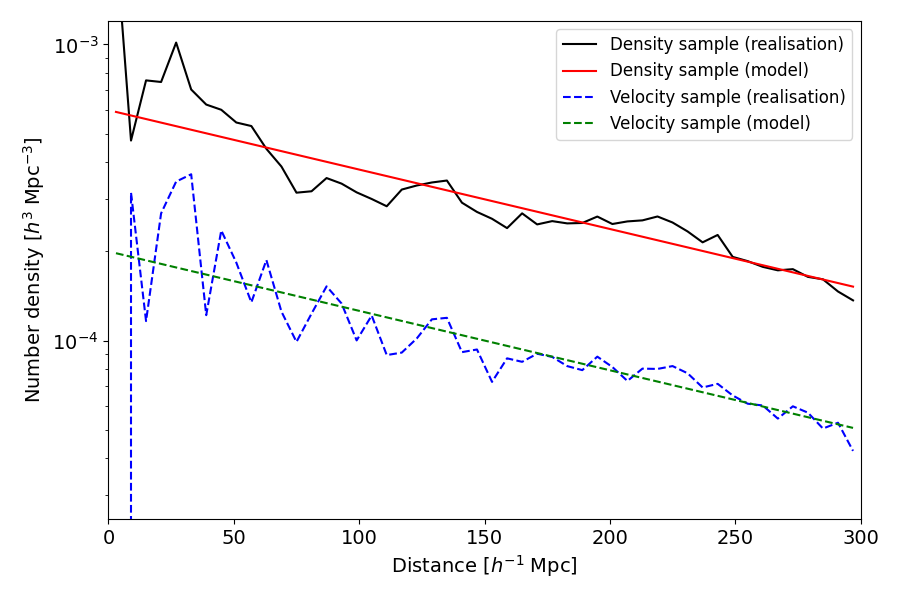}
\end{center}
  \caption{The tracer number density as a function of distance from the observer for the simulated redshift and velocity samples used in this study, comparing the selection function model applied (Eq.\ref{eq:ndens}) to a noisy measurement from an individual realisation.}
  \label{fig:ndens}
\end{figure}

We selected halos within a mass range $12.5 < \log_{10}{(M / h^{-1} M_\odot)} < 13.5$, significantly above the resolution limit, and randomly sub-sampled the density and velocity samples to contain a number density gradient with distance from the observer,
\begin{equation}
n(\vx) = n_{\rm cen}^{(1-x/x_{\rm max})} \cdot n_{\rm edge}^{x/x_{\rm max}} ,
\label{eq:ndens}
\end{equation}
where $n_{\rm cen}$ is the central number density at the observer (with $x=0$), and $n_{\rm edge}$ is the number density at the edge of the sample, where $x = x_{\rm max} = 300 \, h^{-1} {\rm Mpc}$.  For the density sample we chose $(n_{g,{\rm cen}} , n_{g,{\rm edge}}) = (6, 1.5) \times 10^{-4} \, h^3 \, {\rm Mpc}^{-3}$ and for the velocity sample, $(n_{v,{\rm cen}} , n_{v,{\rm edge}}) = (2, 0.5) \times 10^{-4} \, h^3 \, {\rm Mpc}^{-3}$.  These choices produced an average total of $N_g = 2.5 \times 10^4$ density tracers and $N_v = 8.3 \times 10^3$ velocity tracers.  These are arbitrary choices which are representative of existing sample sizes, whilst allowing correlation functions to be swiftly measured.  Fig.\ref{fig:ndens} displays the gradient of tracer density with distance from the observer.  When creating ``observational'' samples we also applied a velocity measurement error which increased with distance from the observer, where the fractional error is $5\%$ of distance,
\begin{equation}
    \epsilon_v(\vx) = 0.05 \, H_0 \, x
\end{equation}
where $H_0 = 100 \, h \, {\rm km} \, {\rm s}^{-1} \, {\rm Mpc}^{-1}$.  This choice of velocity measurement error is representative of the most accurate standard candles such as supernovae \citep{2017MNRAS.464.2517H, 2020MNRAS.498.2703B}, whereas Fundamental Plane measurements from elliptical galaxies would have a fractional error exceeding $20\%$ \citep{2014MNRAS.445.2677S, 2022MNRAS.515..953H}.  We found that different configurations produced broadly similar performance of our covariance model; applying the more accurate noise allows the most accurate test of growth rate recovery.  For ``simulation'' samples using 3D velocities, we measured the velocity noise in each direction as $\sigma_v = 295 \, {\rm km} \, {\rm s}^{-1}$.  We used $N_{\rm mock} = 600$ mock realisations in the following analysis.

\subsection{Correlation function measurements}

We measured the galaxy and velocity auto- and cross-correlation functions of these $N_{\rm mock} = 600$ realisations using standard pair-count estimators.  We used 17 separation bins of width $6 \, h^{-1} \, {\rm Mpc}$ in the range $0 < r < 102 \, h^{-1} \, {\rm Mpc}$.  For the galaxy and inward velocity correlations, we used the pair-count estimators described by \cite{2014JCAP...05..003O} and \cite{2023MNRAS.518.5929L}, based on combinations of weighted sums of pairs from data and random catalogues (where the weights are described below):
\begin{equation}
    \hat{\xi}_{gg}(r) = \left( \frac{N_{Rg}^2}{N_{Dg}^2} \right) \frac{\sum_{Di,Dj} w_{g,i} \, w_{g,j}}{\sum_{Ri,Rj} w_{g,i} \, w_{g,j}} - \frac{2 N_{R,g}}{N_{D,g}} \frac{\sum_{Di,Rj} w_{g,i} \, w_{g,j}}{\sum_{Ri,Rj} w_{g,i} \, w_{g,j}} + 1 ,
\end{equation}
\begin{equation}
    \hat{\xi}_{gv}(r) = \left( \frac{N_{Rg} \, N_{Rv}}{N_{Dg} \, N_{Dv}} \right) \frac{\sum_{Di,Dj} \, w_{g,i} \, w_{v,j} \, v_{r,j}}{\sum_{Ri,Rj} \, w_{g,i} \, w_{v,j}} -  \left( \frac{N_{Rv}}{N_{Dv}} \right) \frac{\sum_{Ri,Dj} \, w_{g,i} \, w_{v,j} \, v_{r,j}}{\sum_{Ri,Rj} \, w_{g,i} \, w_{v,j}} ,
\end{equation}
\begin{equation}
    \hat{\xi}_{vv}(r) = \left( \frac{N_{Rv}^2}{N_{Dv}^2} \right) \frac{\sum_{Di,Dj} w_{v,i} \, w_{v,j} \, v_{r,i} \, v_{r,j}}{\sum_{Ri,Rj} w_{v,i} \, w_{v,j}} .
\end{equation}
In these estimators $w_i$ are the weights of the individual objects, the sums are carried out over pairs in the given separation bin, $N_{Dg}$ and $N_{Dv}$ denote the sum of the weights of the density and velocity tracers, whilst $N_{Rg}$ and $N_{Rv}$ are the corresponding sums over the larger random catalogues.

For the line-of-sight velocity correlations we used the equivalent estimators to Eq.\ref{eq:psi1est}, Eq.\ref{eq:psi2est} and Eq.\ref{eq:psi3est} \citep[see also][]{2021MNRAS.502.2087T},
\begin{equation}
\hat{\psi}_1(r) = \left( \frac{N_{Rv}^2}{N_{Dv}^2} \right) \frac{\sum_{Di,Dj} \, w_{v,i} \, w_{v,j} \, u_i \, u_j \, \cos{\theta_{ij}}}{\sum_{Ri,Rj} \, w_{v,i} \, w_{v,j} \, \cos^2{\theta_{ij}}} ,
\end{equation}
\begin{equation}
\hat{\psi}_2(r) = \left( \frac{N_{Rv}^2}{N_{Dv}^2} \right) \frac{\sum_{Di,Dj} \, w_{v,i} \, w_{v,j} \, u_i \, u_j \, \cos{\theta_i} \, \cos{\theta_j}}{\sum_{Ri,Rj} \, w_{v,i} \, w_{v,j} \, \cos{\theta_{ij}} \, \cos{\theta_i} \, \cos{\theta_j}} ,
\end{equation}
\begin{equation}
\hat{\psi}_3(r) = \left( \frac{N_{Rg} \, N_{Rv}}{N_{Dg} \, N_{Dv}} \right) \frac{\sum_{Di,Dj} \, w_{g,i} \, w_{v,j} \, u_j \, \cos{\theta_j}}{\sum_{Ri,Rj} \, w_{g,i} \, w_{v,j} \, \cos^2{\theta_j}} -  \left( \frac{N_{Rv}}{N_{Dv}} \right) \frac{\sum_{Ri,Dj} \, w_{g,i} \, w_{v,j} \, u_j \, \cos{\theta_j}}{\sum_{Ri,Rj} \, w_{g,i} \, w_{v,j} \, \cos^2{\theta_j}} ,
\end{equation}
where $\theta_i$ is the angle between the line-of-sight vector to the tracer and the separation vector, and $\theta_{ij}$ is the angle between the two line-of-sight vectors.

For the multipoles of $\xi_{gg}$ and $\xi_{gu}$, we measured the correlations in bins of separation, $r$, and the cosine of the angle of the pair to the line-of-sight, $\mu$, and then evaluated,
\begin{equation}
    \hat{\xi}^\ell(s) = (2\ell + 1) \sum_\mu \, \xi(s, \mu) \, L_\ell(\mu) ,
\end{equation}
where we utilised $\xi^0_{gg}$, $\xi^2_{gg}$ and $\xi^1_{gu}$.  We used 20 equal-sized cosine bins in the range $-1 < \mu < +1$.

We assigned the tracers optimal weights.  For the density tracers, we used Feldman-Kaiser-Peacock (FKP) weights \citep{1994ApJ...426...23F},
\begin{equation}
    w_g(\vx) = \frac{1}{1 + n_g(\vx) \, P_{gg}(k_0)} ,
\end{equation}
where we assumed a characteristic galaxy power spectrum amplitude $P_{gg}(k_0) = 10^4 \, h^{-3} \, {\rm Mpc}^3$ (noting that our results are not sensitive to this choice).  For the velocity tracers, we used the analogous optimal weight \citep{2019MNRAS.487.5209H,2021MNRAS.502.2087T},
\begin{equation}
    w_v(\vx) = \frac{1}{\sigma_v^2(\vx) + n_v(\vx) \, P_{vv}(k_0)} ,
\end{equation}
where we assumed a characteristic velocity power spectrum amplitude $P_{vv}(k_0) = 10^{10} \, ({\rm km} \, {\rm s}^{-1})^2 \, h^{-3} \, {\rm Mpc}^3$.

\begin{figure*}
  \includegraphics[width=\columnwidth]{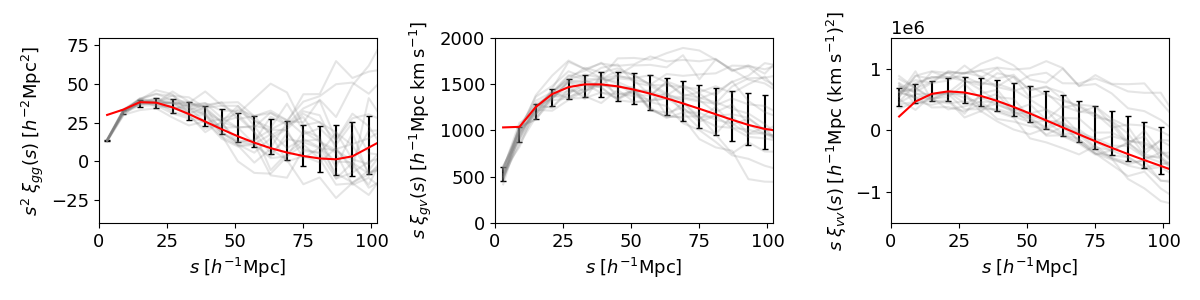}
  \caption{Correlation function measurements for the ``simulation'' data vector in which we utilise 3D velocities and analyse the three correlation functions $[ \xi_{gg} , \xi_{gv} , \xi_{vv} ]$.  The black error bars illustrate the standard deviation of the measured correlations around the mock mean, and the faint, grey solid lines indicate the measurements of the first 20 individual realisations.  The red solid lines represent the fiducial models.}
  \label{fig:corr_norsd}
\end{figure*}

\begin{figure*}
  \includegraphics[width=\columnwidth]{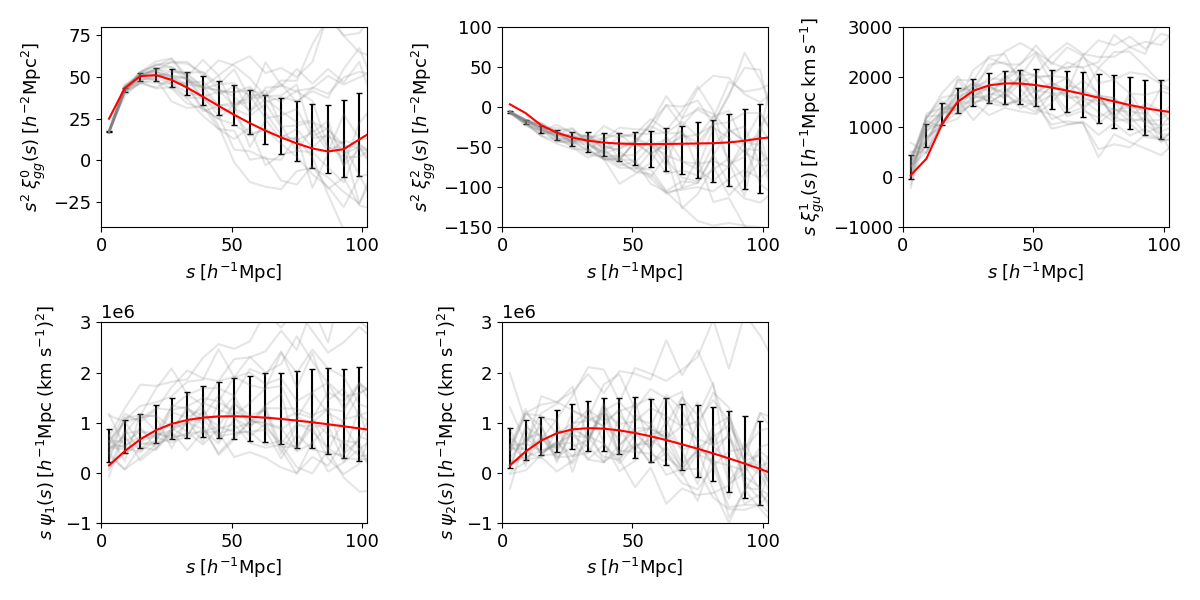}
  \caption{Correlation function measurements for the ``observation'' data vector in which we utilise line-of-sight velocities and redshift-space multipoles, and analyse the five correlation functions $[ \xi^0_{gg} , \xi^2_{gg} , \xi^1_{gu} , \psi_1, \psi_2 ]$.  The black error bars illustrate the standard deviation of the measured correlations around the mock mean, and the faint, grey solid lines indicate the measurements of the first 20 individual realisations.  The red solid lines represent the fiducial models.}
  \label{fig:corr_rsd}
\end{figure*}

We grouped the correlation functions into two separate data vectors: the 3D velocity ``simulation'' case involving the three correlations $[ \xi_{gg} , \xi_{gv} , \xi_{vv} ]$ without redshift-space distortions, and the line-of-sight velocity ``observation'' case involving the five correlation functions $[ \xi^0_{gg} , \xi^2_{gg} , \xi^1_{gu} , \psi_1, \psi_2 ]$ including redshift-space distortions.  The mock mean, standard deviation and fiducial models are displayed for the ``simulation'' case in Fig.\ref{fig:corr_norsd}, and for the ``observation'' case in Fig.\ref{fig:corr_rsd}.  We generated the fiducial models by evaluating Eq.\ref{eq:ggcorr2} (for $\xi_{gg}$), Eq.\ref{eq:gvcorr} (for $\xi_{gv}$), Eq.\ref{eq:vvcorr} (for $\xi_{vv}$), Eq.\ref{eq:ggcorrmult} (for $\xi^\ell_{gg}$), Eq.\ref{eq:gucorrmult} (for $\xi^\ell_{gu}$), Eq.\ref{eq:psi1est} (for $\psi_1$) and Eq.\ref{eq:psi2est} (for $\psi_2$).  We created the model matter power spectrum $P_m(k)$ using the original ``halofit'' algorithm \citep{2003MNRAS.341.1311S} within the CAMB software \citep{2000ApJ...538..473L}, assuming a fiducial galaxy bias factor $b_g = 1.04$, which we fixed through comparison of the measured and theoretical galaxy power spectra.  When integrating models over $k$, we assumed a lower limit given by the box size $k_{\rm min} = \frac{2\pi}{L} \approx 0.01 \, h \, {\rm Mpc}^{-1}$.  The agreement between the correlation function models and measurements worsens on scales $s < 20 \, h^{-1}$ Mpc, where the linear-theory modelling becomes less accurate.

\subsection{Covariance comparison}

We evaluated the analytical covariance for the different components of the ``simulation'' and ``observation'' data vectors, using the approximations defined in Sec.\ref{sec:anacov}.  We first determined the various averages of the different selection function and the noise fields over the embedding cube (as defined by Eq.\ref{eq:selave}), and then evaluated the integrals listed in Appendix \ref{sec:allthecovs}.  We increased the accuracy of the determination by averaging the Bessel functions within these integrals over each separation bin.

\begin{figure*}
  \includegraphics[width=\columnwidth]{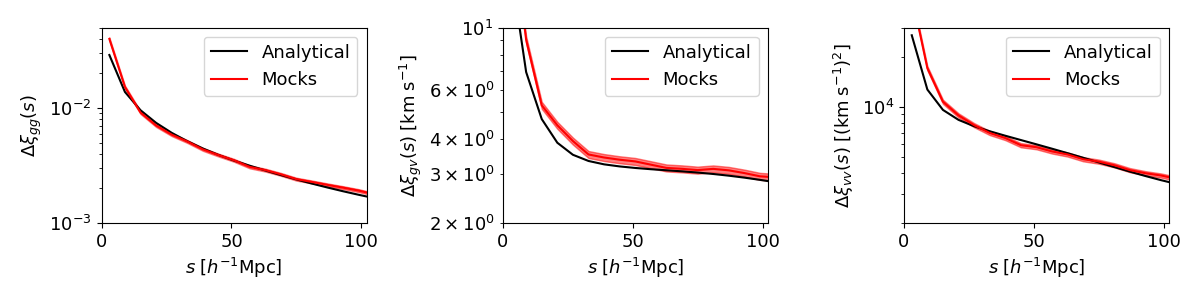}
  \caption{Comparison of the standard deviation across the mocks of the correlation function measurements in each separation bin (the red band, with width indicating the error in the standard deviation), with the prediction of the analytical covariance (solid black line).  This figure uses the ``simulation'' data vector, in which we utilise 3D velocities and analyse the three correlation functions $[ \xi_{gg} , \xi_{gv} , \xi_{vv} ]$.}
  \label{fig:correrr_norsd}
\end{figure*}

\begin{figure*}
  \includegraphics[width=\columnwidth]{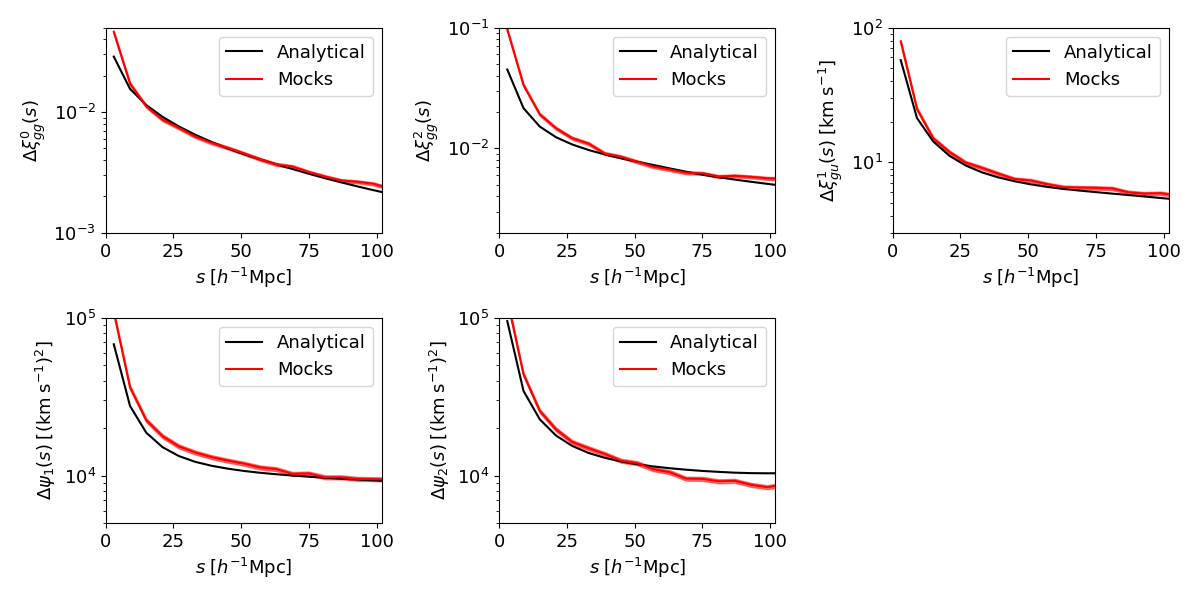}
  \caption{Comparison of the standard deviation across the mocks of the correlation function measurements in each separation bin (the red band, with width indicating the error in the standard deviation), with the prediction of the analytical covariance (solid black line).  This figure uses the ``observation'' data vector, in which we utilise line-of-sight velocities and redshift-space multipoles, and analyse the five correlation functions $[ \xi^0_{gg} , \xi^2_{gg} , \xi^1_{gu} , \psi_1, \psi_2 ]$.}
  \label{fig:correrr_rsd}
\end{figure*}

Figs. \ref{fig:correrr_norsd} and \ref{fig:correrr_rsd} illustrate the comparison between the predicted analytical error in each separation bin, and the standard deviation of the mock measurements, for the individual correlation functions within the ``simulation'' and ``observation'' case, respectively.  The mock measurements are displayed as a band, where the width of the band indicates the ``error in the error'' appropriate for Gaussian statistics, $\Delta \sigma \approx \sigma/\sqrt{2 \, N_{\rm mock}}$.  We find that, generally speaking, the analytical covariance provides an accurate model for the dispersion between mocks, with the principal deviations at small scales.  The average difference between the analytical and mock error in each separation bin, across the different statistics, is below $5\%$.

\begin{figure*}
  \includegraphics[width=\columnwidth]{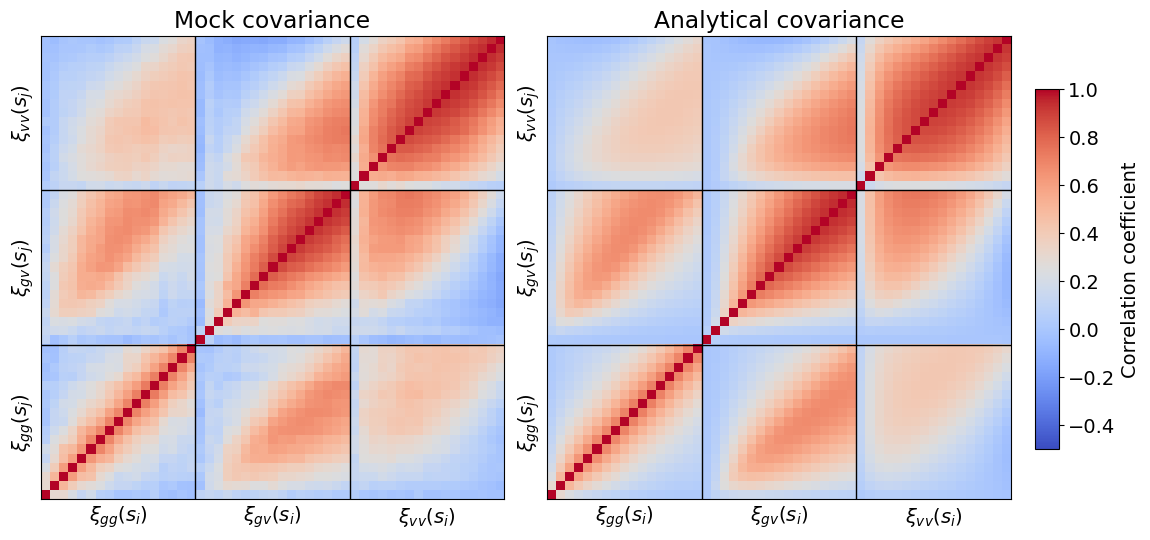}
  \caption{Comparison of the correlation coefficients of the joint covariance, $r_{ij} = C_{ij}/\sqrt{(C_{ii} \, C_{jj})}$, for the ``simulation'' case in which we utilise 3D velocities and the data vector is composed of the three correlation functions $[ \xi_{gg} , \xi_{gv} , \xi_{vv} ]$.  The left-hand and right-hand panels display the correlation coefficient of the full mock covariance, and analytical covariance, respectively.  Each individual sub-panel is composed of the 17 separation bins in the range $0 < s < 102 \, h^{-1} \, {\rm Mpc}$.  The range of the colour map is $-0.5 < r < 1$, as indicated by the colour bar.}
  \label{fig:cov_norsd}
\end{figure*}

\begin{figure*}
  \includegraphics[width=\columnwidth]{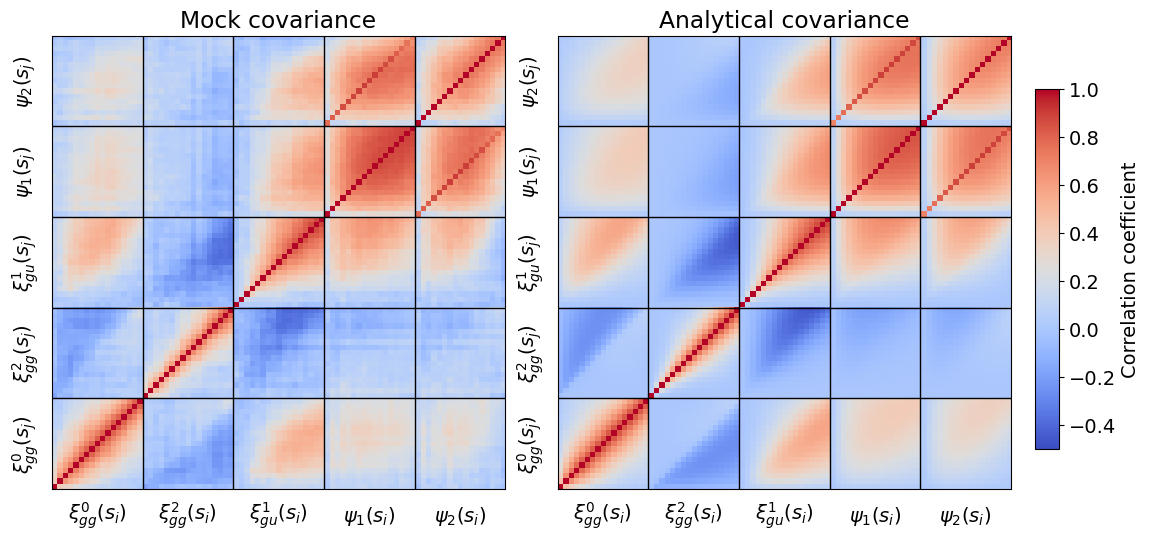}
  \caption{Comparison of the correlation coefficients of the joint covariance, $r_{ij} = C_{ij}/\sqrt{(C_{ii} \, C_{jj})}$, for the ``observation'' case in which we utilise line-of-sight velocities and redshift-space multipoles, and the data vector is composed of the five correlation functions $[ \xi^0_{gg} , \xi^2_{gg} , \xi^1_{gu} , \psi_1, \psi_2 ]$.  The left-hand and right-hand panels display the correlation coefficient of the full mock covariance, and analytical covariance, respectively.  Each individual sub-panel is composed of the 17 separation bins in the range $0 < s < 102 \, h^{-1} \, {\rm Mpc}$.  The range of the colour map is $-0.5 < r < 1$, as indicated by the colour bar.}
  \label{fig:cov_rsd}
\end{figure*}

Figs. \ref{fig:cov_norsd} and \ref{fig:cov_rsd} display the correlation coefficients of the full covariance matrix derived from the mock measurements, compared to the analytical calculation, spanning the different correlation functions.  Fig. \ref{fig:cov_norsd} uses the ``simulation'' data vector including the three correlation functions $[ \xi_{gg} , \xi_{gv} , \xi_{vv} ]$, and Fig. \ref{fig:cov_rsd} uses the ``observation'' data vector encompassing the five correlation functions $[ \xi^0_{gg} , \xi^2_{gg} , \xi^1_{gu} , \psi_1, \psi_2 ]$.  We can see that the ``broad features'' of the correlation patterns are successfully reproduced in each case.

We quantified the difference between the analytical and mock covariance matrices using two initial methods.  First, we considered the impact on the errors in the model parameters $p_i$ by computing the associated Fisher matrices $F_{ij}$,
\begin{equation}
  F_{ij} = \sum_{k,l} \frac{\partial m_k}{\partial p_i} \, C^{-1}_{kl} \, \frac{\partial m_l}{\partial p_j} ,
\end{equation}
where $\mathbf{m}$ is the model vector, $\mathbf{C}$ is the covariance, and the forecast parameter errors are given by $\Delta p_i = \sqrt{\left( \mathbf{F}^{-1} \right)_{ii}}$.  We considered a parameter set $p_i = (f,b)$ for the ``simulation'' dataset, and $p_i = (f,b,\sigma_v)$ for the ``observation'' dataset (where the model is specified more fully in Sec.\ref{sec:growthfit}).  The Fisher forecast parameter errors of the two covariances agreed within $10\%$ for both correlation function sets.  Second, following \cite{2021A&A...646A.129J}, we assessed any systematic offset between the best-fitting parameters for each covariance by sampling many noisy data vectors $d_i$,
\begin{equation}
  d_i = m_i + \sum_j \, L_{ij} \, z_j ,
\end{equation}
where $\mathbf{L}$ is a triangular matrix obtained from the Cholesky decomposition of the covariance matrix, $\mathbf{C} = \mathbf{L} \, \mathbf{L}^T$, and $\mathbf{z}$ is a vector of random variables drawn from a Gaussian distribution with zero mean and unit variance.  We generated noisy data vectors from the fiducial model for both the mock and analytical covariance, fit these vectors for the growth rate and galaxy bias parameters, and considered the offset in the best-fitting growth rate parameters.  This offset was less than $10\%$ of the statistical error in the fit.  In the following subsection, we consider a more comprehensive comparison of the two covariances using full model fits to the mock correlation functions.

\subsection{Growth rate fits}
\label{sec:growthfit}

We can use the analytical covariance matrices derived in Sec.\ref{sec:anacov} and the full mock covariance matrices described in Sec.\ref{sec:mocks}, all shown in Fig.\ref{fig:cov_norsd} and Fig.\ref{fig:cov_rsd}, to produce fits for the growth rate $f$ by means of a $\chi^2$ minimisation procedure. We reproduce the analyses of \cite{2021MNRAS.502.2087T} and \cite{2023MNRAS.518.2436T}, using the former for our 3D velocity ``simulation" case neglecting redshift-space distortions and using the latter for our line-of-sight velocity ``observation" case that includes the effects of redshift-space distortions, fitting to separations greater than $20 \, h^{-1}$ Mpc where our linear-theory models are applicable.

\subsubsection{Simulation case}

\begin{figure*}
    \includegraphics[width=\columnwidth]{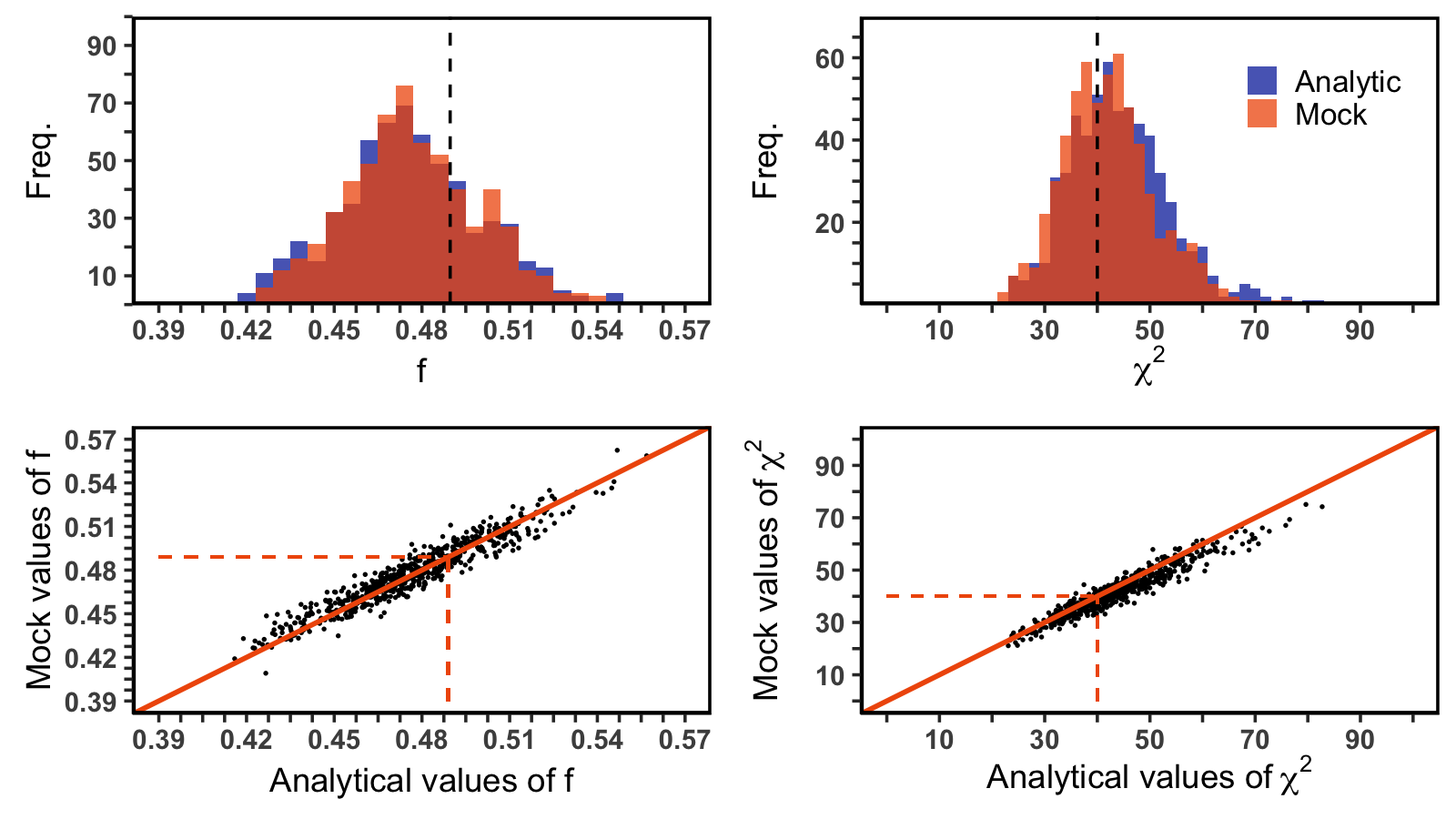}
    \caption{Comparison between fits for the growth rate $f$ made using the full mock covariance and the analytical covariance, utilising 3D velocities and where the data vector is composed of the three correlation functions $[ \xi_{gg} , \xi_{gv} , \xi_{vv} ]$. The panels in the top row display the distribution of best-fitting parameters obtained from all 600 {\sc COLA} mocks (left) and the corresponding minimum $\chi^2$ values (right), the results obtained with the full mock covariance are in orange, those from the analytic covariance are in blue and the vertical dashed line depicts the fiducial value of $f$ (left) and the number of degrees of freedom in the $\chi^2$ fit (right). The panels in the bottom row depict the same data but as a direct comparison between the analytical covariance results on the x axis, and full mock covariance results on the y axis. A line of equivalence, $y = x$, is shown in orange on both panels and the dashed lines represent the same values as the dashed orange lines in the upper panels.}
    \label{fig:3corr}
\end{figure*}

Following \cite{2021MNRAS.502.2087T}, we use the three correlations $[ \xi_{gg} , \xi_{gv} , \xi_{vv} ]$ to produce a measurement of $f$. As can be seen from the fiducial models of these functions (Eq.\ref{eq:ggcorr2}, Eq.\ref{eq:gvcorr}, and Eq.\ref{eq:vvcorr} respectively) and the definitions of the corresponding power spectra $[ P_{gg}(k) , P_{gv}(k) , P_{vv}(k) ]$ given in Sec.\ref{sec:corrstat}, these correlation functions are differently dependent on the growth rate $f$ and the linear galaxy bias $b$. By fitting $f$ and $b$ to all three models simultaneously in a self-consistent manner we can jointly constrain both parameters. Given the assumption that the full 3D velocity information is known, we simply calculate Eq.\ref{eq:ggcorr2}, Eq.\ref{eq:gvcorr}, and Eq.\ref{eq:vvcorr} for each of the 600 {\sc COLA} mocks and use a $\chi^2$ minimisation to determine the best-fitting values of $f$ for each mock as determined from the minimum $\chi^2$ value -- where $\chi^2$ is calculated using the following equation:
\begin{equation}
    \chi^2(f, b) = \sum_{i,j = 1}^N\,(A_d(i) - A_m(i; f, b))\,C_{ij}^{-1}\,(A_d(j) - A_m(j; f, b)) ,
    \label{eq:chisq}
\end{equation}
where $A_d$ are the data vectors composed of the three correlation functions calculated from all 600 {\sc COLA} mocks, $A_m$ are the fiducial models of each correlation function, and $C^{-1}$ is the inverted covariance matrix. To compare the accuracy of the analytical covariance in relation to the full mock covariance, we evaluate Eq.\ref{eq:chisq} using both matrices in place of $C$ and contrast the final results. This comparison is shown in Fig.\ref{fig:3corr}. We can see from the figure that the results obtained from the analytical analysis and those from the mock analysis are consistent, and both recover the fiducial value of the growth rate as indicated by the dashed vertical line. Taking all 600 {\sc COLA} mocks into account, the analytical ensemble mean value of the growth rate is $f_{\rm ana} = 0.476 \pm 0.024$ with an average reduced $\chi^2$ value of $\chi^2_{\nu} = 1.10$, and the mock ensemble mean value is $f_{\rm mock} = 0.477 \pm 0.023$ with an average reduced $\chi^2$ value of $\chi^2_{\nu} = 1.05$. Taking the difference between the best-fitting value of $f_{\rm ana}$ and $f_{\rm mock}$ for each of the 600 mocks, we find the standard deviation across the full ensemble to be $\sigma_{\rm diff} = 0.004$. By comparing the scatter in the difference between our analytical and mock results with the scatter in $f_{\rm ana}$ or $f_{\rm mock}$, we find that the systematic error inherent to choosing either the analytical or full mock covariance is negligible and thus the analytical covariance matrix can reliably be used in place of a full mock covariance matrix.

\subsubsection{Observation case}

Similarly to the simulation case, now following the methodology of \cite{2023MNRAS.518.2436T}, we use the five correlation functions including redshift-space distortions $[ \xi^0_{gg} , \xi^2_{gg} , \xi^1_{gu} , \psi_1, \psi_2 ]$  to produce a measurement of $f$. Using fiducial models defined by Eq.\ref{eq:ggcorrmult}, Eq.\ref{eq:gucorrmult}, Eq.\ref{eq:psi1corr} and Eq.\ref{eq:psi2corr} for each function, respectively, and their corresponding radial-velocity-dependent estimators given by Eq.\ref{eq:xiggest}, Eq.\ref{eq:xiguest}, Eq.\ref{eq:psi1est} and Eq.\ref{eq:psi2est} we can again produce best-fitting measurements of the growth rate $f$. We implement a similar $\chi^2$ minimisation procedure, using the equation:
\begin{equation}
    \chi^2(\beta, b, \sigma_v) = \sum_{i,j = 1}^N\,(A_d(i) - A_m(i; \beta, b, \sigma_v))\,C_{ij}^{-1}\,(A_d(j) - A_m(j; \beta, b, \sigma_v))
    \label{eq:chisq_rsd}
\end{equation}
where $A_d$ are the data vectors composed of the five correlation function estimators calculated from all 600 {\sc COLA} mocks and $A_m$ are the fiducial models of the five correlation functions. Given that these correlation functions incorporate redshift-space distortions we cast the multipole models in terms of the linear redshift distortion parameter $\beta \equiv f/b$ and a non-linear velocity dispersion parameter $\sigma_v$ with units of km s$^{-1}$ \citep[see][for the exact form of these models]{2023MNRAS.518.2436T}. Considering the best-fitting values of $f$ from all 600 {\sc COLA} mocks, the analytical ensemble mean value of the growth rate is $f_{\rm ana} = 0.496 \pm 0.063$ with an average reduced $\chi^2$ value of $\chi^2_{\nu} = 1.09$, and the mock ensemble mean value is $f_{\rm mock} = 0.485 \pm 0.049$ with an average reduced $\chi^2$ value of $\chi^2_{\nu} = 1.01$. Again measuring the standard deviation in the differences between the best-fitting values of $f_{\rm ana}$ and $f_{\rm mock}$ for the 600 mocks, we find $\sigma_{\rm diff} = 0.029$. Echoing the results seen in the three correlation function ``simulation" case, the scatter due to the choice of covariance matrix is significantly less than the error in the ensemble mean measurement of $f_{\rm ana}$ or $f_{\rm mock}$.

\begin{figure*}
    \includegraphics[width=\columnwidth]{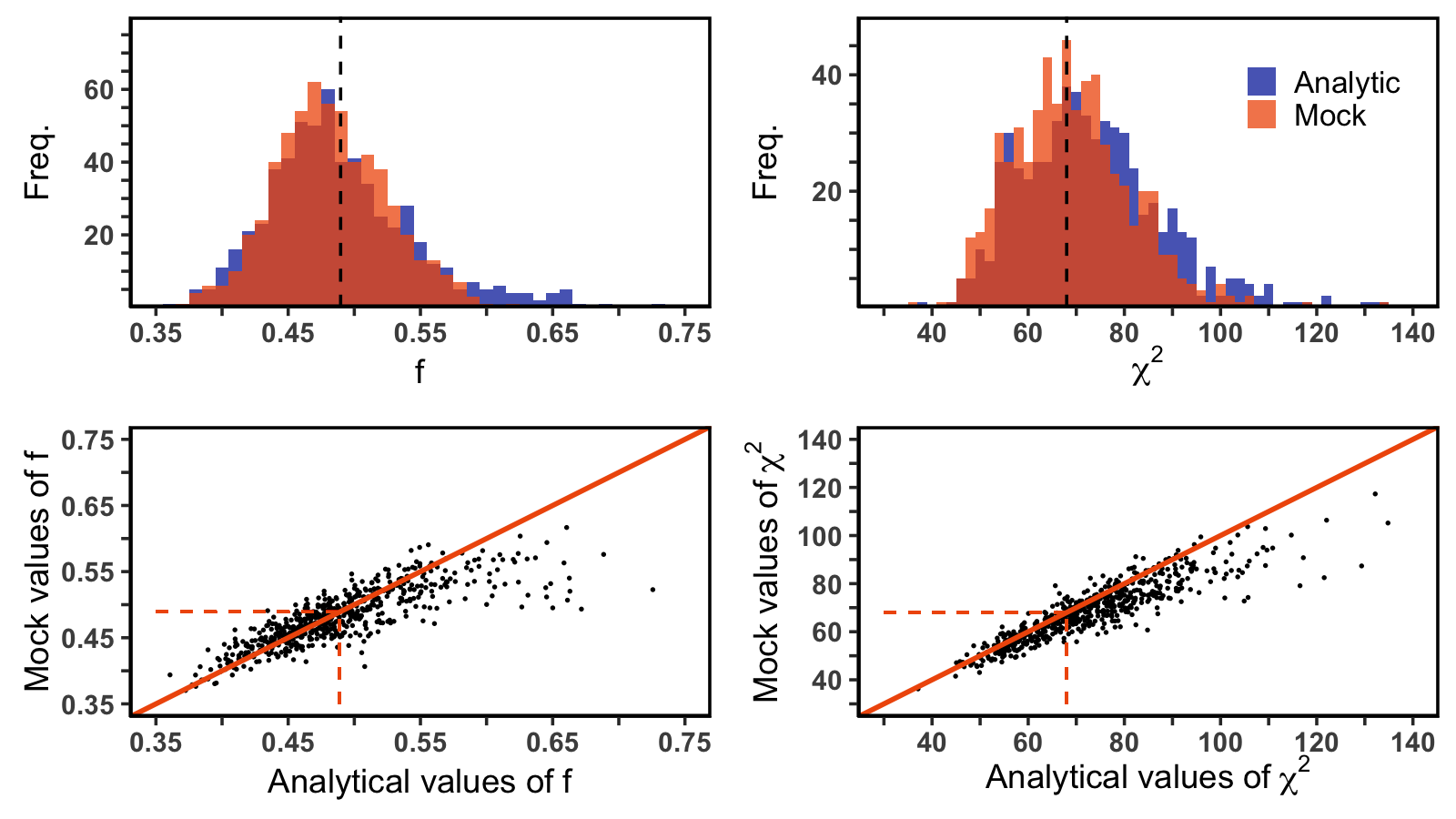}
    \caption{Comparison between fits for the growth rate $f$ made using the full mock covariance and the analytical covariance, utilising line-of-sight velocities and where the data vector is composed of the five correlation functions $[ \xi^0_{gg} , \xi^2_{gg} , \xi^1_{gu} , \psi_1, \psi_2 ]$. The panels in the top row display the distribution of $f$ results obtained from all 600 {\sc COLA} mocks (left) and the corresponding minimum $\chi^2$ values (right), the results obtained with the full mock covariance are in orange, those from the analytic covariance are in blue and the vertical dashed line depicts the fiducial value of $f$ (left) and the number of degrees of freedom in the $\chi^2$ fit (right). The panels in the bottom row depict the same data but as a direct comparison between the analytical covariance results on the x axis, and full mock covariance results on the y axis. A line of equivalence, $y = x$, is shown in orange on both panels and the dashed orange lines represent the same values as the dashed lines in the upper panels.}
    \label{fig:5corr}
\end{figure*}

\section{Conclusions}
\label{sec:conc}

Large samples of galaxies, tracing the matter density field and measuring direct peculiar velocities, can be summarised via galaxy and velocity auto- and cross-correlation functions, which form a convenient point of comparison with theoretical models.  The upcoming generation of large cosmological surveys will provide velocity samples containing hundreds of thousands of galaxies, promising new and accurate insights into gravitational physics on cosmological scales.  A potential obstacle in such analyses is determining the covariance of these statistics, which is an important ingredient in Bayesian likelihood analyses.   In this paper we have presented new calculations of the analytical covariance between different types of velocity correlation functions and scales.  We considered both the ``simulation'' case, in which we analyse 3D velocities, and the ``observation'' case, in which we utilise line-of-sight velocities and redshift-space multipoles.  The statistical model used in our covariance includes the survey selection function, observational noise, curved-sky effects and redshift-space distortions.  We compared our analytical covariance determinations with equivalent covariances estimated using a large suite of cosmological simulations, demonstrating that the analytical covariance is a good representation of the dispersion across the simulations.  We also compared the performance of both covariances in fits for the growth rate of cosmic structure, the key cosmological parameter which may be tested by such datasets, finding that the respective growth rate determinations are comparable.  Our study demonstrates that the analytical covariance can be a useful tool in the analysis of large velocity samples, in cases where sufficient numerical simulations are unavailable.

\section*{Acknowledgements}

We thank the anonymous referee for useful comments on the manuscript.  We acknowledge financial support received through Australian Research Council Discovery Project DP220101610.

\section*{Data Availability}

The data underlying this article will be shared on reasonable request to the corresponding author.

\bibliographystyle{mnras}
\bibliography{velocity_covariance}

\begin{thebibliography}{}
\makeatletter
\relax
\def\mn@urlcharsother{\let\do\@makeother \do\$\do\&\do\#\do\^\do\_\do\%\do\~}
\def\mn@doi{\begingroup\mn@urlcharsother \@ifnextchar [ {\mn@doi@}
  {\mn@doi@[]}}
\def\mn@doi@[#1]#2{\def\@tempa{#1}\ifx\@tempa\@empty \href
  {http://dx.doi.org/#2} {doi:#2}\else \href {http://dx.doi.org/#2} {#1}\fi
  \endgroup}
\def\mn@eprint#1#2{\mn@eprint@#1:#2::\@nil}
\def\mn@eprint@arXiv#1{\href {http://arxiv.org/abs/#1} {{\tt arXiv:#1}}}
\def\mn@eprint@dblp#1{\href {http://dblp.uni-trier.de/rec/bibtex/#1.xml}
  {dblp:#1}}
\def\mn@eprint@#1:#2:#3:#4\@nil{\def\@tempa {#1}\def\@tempb {#2}\def\@tempc
  {#3}\ifx \@tempc \@empty \let \@tempc \@tempb \let \@tempb \@tempa \fi \ifx
  \@tempb \@empty \def\@tempb {arXiv}\fi \@ifundefined
  {mn@eprint@\@tempb}{\@tempb:\@tempc}{\expandafter \expandafter \csname
  mn@eprint@\@tempb\endcsname \expandafter{\@tempc}}}

\bibitem[\protect\citeauthoryear{{Adams} \& {Blake}}{{Adams} \&
  {Blake}}{2017}]{2017MNRAS.471..839A}
{Adams} C.,  {Blake} C.,  2017, \mn@doi [\mnras] {10.1093/mnras/stx1529}, \href
  {https://ui.adsabs.harvard.edu/abs/2017MNRAS.471..839A} {471, 839}

\bibitem[\protect\citeauthoryear{{Adams} \& {Blake}}{{Adams} \&
  {Blake}}{2020}]{2020MNRAS.494.3275A}
{Adams} C.,  {Blake} C.,  2020, \mn@doi [\mnras] {10.1093/mnras/staa845}, \href
  {https://ui.adsabs.harvard.edu/abs/2020MNRAS.494.3275A} {494, 3275}

\bibitem[\protect\citeauthoryear{{Blake}}{{Blake}}{2019}]{2019MNRAS.489..153B}
{Blake} C.,  2019, \mn@doi [\mnras] {10.1093/mnras/stz2145}, \href
  {https://ui.adsabs.harvard.edu/abs/2019MNRAS.489..153B} {489, 153}

\bibitem[\protect\citeauthoryear{{Blake}, {Carter}  \& {Koda}}{{Blake}
  et~al.}{2018}]{2018MNRAS.479.5168B}
{Blake} C.,  {Carter} P.,   {Koda} J.,  2018, \mn@doi [\mnras]
  {10.1093/mnras/sty1814}, \href
  {https://ui.adsabs.harvard.edu/abs/2018MNRAS.479.5168B} {479, 5168}

\bibitem[\protect\citeauthoryear{{Boruah}, {Hudson}  \& {Lavaux}}{{Boruah}
  et~al.}{2020}]{2020MNRAS.498.2703B}
{Boruah} S.~S.,  {Hudson} M.~J.,   {Lavaux} G.,  2020, \mn@doi [\mnras]
  {10.1093/mnras/staa2485}, \href
  {https://ui.adsabs.harvard.edu/abs/2020MNRAS.498.2703B} {498, 2703}

\bibitem[\protect\citeauthoryear{{Burkey} \& {Taylor}}{{Burkey} \&
  {Taylor}}{2004}]{2004MNRAS.347..255B}
{Burkey} D.,  {Taylor} A.~N.,  2004, \mn@doi [\mnras]
  {10.1111/j.1365-2966.2004.07192.x}, \href
  {https://ui.adsabs.harvard.edu/abs/2004MNRAS.347..255B} {347, 255}

\bibitem[\protect\citeauthoryear{{Carrick}, {Turnbull}, {Lavaux}  \&
  {Hudson}}{{Carrick} et~al.}{2015}]{2015MNRAS.450..317C}
{Carrick} J.,  {Turnbull} S.~J.,  {Lavaux} G.,   {Hudson} M.~J.,  2015, \mn@doi
  [\mnras] {10.1093/mnras/stv547}, \href
  {https://ui.adsabs.harvard.edu/abs/2015MNRAS.450..317C} {450, 317}

\bibitem[\protect\citeauthoryear{{Courtois} et~al.,}{{Courtois}
  et~al.}{2023a}]{2023MNRAS.519.4589C}
{Courtois} H.~M.,  et~al., 2023a, \mn@doi [\mnras] {10.1093/mnras/stac3246},
  \href {https://ui.adsabs.harvard.edu/abs/2023MNRAS.519.4589C} {519, 4589}

\bibitem[\protect\citeauthoryear{{Courtois}, {Dupuy}, {Guinet}, {Baulieu},
  {Ruppin}  \& {Brenas}}{{Courtois} et~al.}{2023b}]{2023A&A...670L..15C}
{Courtois} H.~M.,  {Dupuy} A.,  {Guinet} D.,  {Baulieu} G.,  {Ruppin} F.,
  {Brenas} P.,  2023b, \mn@doi [\aap] {10.1051/0004-6361/202245331}, \href
  {https://ui.adsabs.harvard.edu/abs/2023A&A...670L..15C} {670, L15}

\bibitem[\protect\citeauthoryear{{Dam}, {Bolejko}  \& {Lewis}}{{Dam}
  et~al.}{2021}]{2021JCAP...09..018D}
{Dam} L.,  {Bolejko} K.,   {Lewis} G.~F.,  2021, \mn@doi [\jcap]
  {10.1088/1475-7516/2021/09/018}, \href
  {https://ui.adsabs.harvard.edu/abs/2021JCAP...09..018D} {2021, 018}

\bibitem[\protect\citeauthoryear{{Dupuy}, {Courtois}  \& {Kubik}}{{Dupuy}
  et~al.}{2019}]{2019MNRAS.486..440D}
{Dupuy} A.,  {Courtois} H.~M.,   {Kubik} B.,  2019, \mn@doi [\mnras]
  {10.1093/mnras/stz901}, \href
  {https://ui.adsabs.harvard.edu/abs/2019MNRAS.486..440D} {486, 440}

\bibitem[\protect\citeauthoryear{{Favole}, {Granett}, {Silva Lafaurie}  \&
  {Sapone}}{{Favole} et~al.}{2021}]{2021MNRAS.505.5833F}
{Favole} G.,  {Granett} B.~R.,  {Silva Lafaurie} J.,   {Sapone} D.,  2021,
  \mn@doi [\mnras] {10.1093/mnras/stab1720}, \href
  {https://ui.adsabs.harvard.edu/abs/2021MNRAS.505.5833F} {505, 5833}

\bibitem[\protect\citeauthoryear{{Feldman}, {Kaiser}  \& {Peacock}}{{Feldman}
  et~al.}{1994}]{1994ApJ...426...23F}
{Feldman} H.~A.,  {Kaiser} N.,   {Peacock} J.~A.,  1994, \mn@doi [\apj]
  {10.1086/174036}, \href
  {https://ui.adsabs.harvard.edu/abs/1994ApJ...426...23F} {426, 23}

\bibitem[\protect\citeauthoryear{{Fisher}}{{Fisher}}{1995}]{1995ApJ...448..494F}
{Fisher} K.~B.,  1995, \mn@doi [\apj] {10.1086/175980}, \href
  {https://ui.adsabs.harvard.edu/abs/1995ApJ...448..494F} {448, 494}

\bibitem[\protect\citeauthoryear{{Friedrich} et~al.,}{{Friedrich}
  et~al.}{2021}]{2021MNRAS.508.3125F}
{Friedrich} O.,  et~al., 2021, \mn@doi [\mnras] {10.1093/mnras/stab2384}, \href
  {https://ui.adsabs.harvard.edu/abs/2021MNRAS.508.3125F} {508, 3125}

\bibitem[\protect\citeauthoryear{{Gorski}}{{Gorski}}{1988}]{1988ApJ...332L...7G}
{Gorski} K.,  1988, \mn@doi [\apjl] {10.1086/185255}, \href
  {https://ui.adsabs.harvard.edu/abs/1988ApJ...332L...7G} {332, L7}

\bibitem[\protect\citeauthoryear{{Gorski}, {Davis}, {Strauss}, {White}  \&
  {Yahil}}{{Gorski} et~al.}{1989}]{1989ApJ...344....1G}
{Gorski} K.~M.,  {Davis} M.,  {Strauss} M.~A.,  {White} S. D.~M.,   {Yahil} A.,
   1989, \mn@doi [\apj] {10.1086/167771}, \href
  {https://ui.adsabs.harvard.edu/abs/1989ApJ...344....1G} {344, 1}

\bibitem[\protect\citeauthoryear{{Grieb}, {S{\'a}nchez}, {Salazar-Albornoz}  \&
  {Dalla Vecchia}}{{Grieb} et~al.}{2016}]{2016MNRAS.457.1577G}
{Grieb} J.~N.,  {S{\'a}nchez} A.~G.,  {Salazar-Albornoz} S.,   {Dalla Vecchia}
  C.,  2016, \mn@doi [\mnras] {10.1093/mnras/stw065}, \href
  {https://ui.adsabs.harvard.edu/abs/2016MNRAS.457.1577G} {457, 1577}

\bibitem[\protect\citeauthoryear{{Hartlap}, {Simon}  \& {Schneider}}{{Hartlap}
  et~al.}{2007}]{2007A&A...464..399H}
{Hartlap} J.,  {Simon} P.,   {Schneider} P.,  2007, \mn@doi [\aap]
  {10.1051/0004-6361:20066170}, \href
  {https://ui.adsabs.harvard.edu/abs/2007A&A...464..399H} {464, 399}

\bibitem[\protect\citeauthoryear{{Hou}, {Cahn}, {Philcox}  \& {Slepian}}{{Hou}
  et~al.}{2022}]{2022PhRvD.106d3515H}
{Hou} J.,  {Cahn} R.~N.,  {Philcox} O. H.~E.,   {Slepian} Z.,  2022, \mn@doi
  [\prd] {10.1103/PhysRevD.106.043515}, \href
  {https://ui.adsabs.harvard.edu/abs/2022PhRvD.106d3515H} {106, 043515}

\bibitem[\protect\citeauthoryear{{Howlett}}{{Howlett}}{2019}]{2019MNRAS.487.5209H}
{Howlett} C.,  2019, \mn@doi [\mnras] {10.1093/mnras/stz1403}, \href
  {https://ui.adsabs.harvard.edu/abs/2019MNRAS.487.5209H} {487, 5209}

\bibitem[\protect\citeauthoryear{{Howlett}, {Staveley-Smith}  \&
  {Blake}}{{Howlett} et~al.}{2017a}]{2017MNRAS.464.2517H}
{Howlett} C.,  {Staveley-Smith} L.,   {Blake} C.,  2017a, \mn@doi [\mnras]
  {10.1093/mnras/stw2466}, \href
  {https://ui.adsabs.harvard.edu/abs/2017MNRAS.464.2517H} {464, 2517}

\bibitem[\protect\citeauthoryear{{Howlett}, {Robotham}, {Lagos}  \&
  {Kim}}{{Howlett} et~al.}{2017b}]{2017ApJ...847..128H}
{Howlett} C.,  {Robotham} A. S.~G.,  {Lagos} C. D.~P.,   {Kim} A.~G.,  2017b,
  \mn@doi [\apj] {10.3847/1538-4357/aa88c8}, \href
  {https://ui.adsabs.harvard.edu/abs/2017ApJ...847..128H} {847, 128}

\bibitem[\protect\citeauthoryear{{Howlett}, {Said}, {Lucey}, {Colless}, {Qin},
  {Lai}, {Tully}  \& {Davis}}{{Howlett} et~al.}{2022}]{2022MNRAS.515..953H}
{Howlett} C.,  {Said} K.,  {Lucey} J.~R.,  {Colless} M.,  {Qin} F.,  {Lai} Y.,
  {Tully} R.~B.,   {Davis} T.~M.,  2022, \mn@doi [\mnras]
  {10.1093/mnras/stac1681}, \href
  {https://ui.adsabs.harvard.edu/abs/2022MNRAS.515..953H} {515, 953}

\bibitem[\protect\citeauthoryear{{Huterer}, {Shafer}, {Scolnic}  \&
  {Schmidt}}{{Huterer} et~al.}{2017}]{2017JCAP...05..015H}
{Huterer} D.,  {Shafer} D.~L.,  {Scolnic} D.~M.,   {Schmidt} F.,  2017, \mn@doi
  [\jcap] {10.1088/1475-7516/2017/05/015}, \href
  {https://ui.adsabs.harvard.edu/abs/2017JCAP...05..015H} {2017, 015}

\bibitem[\protect\citeauthoryear{{Joachimi} et~al.,}{{Joachimi}
  et~al.}{2021}]{2021A&A...646A.129J}
{Joachimi} B.,  et~al., 2021, \mn@doi [\aap] {10.1051/0004-6361/202038831},
  \href {https://ui.adsabs.harvard.edu/abs/2021A&A...646A.129J} {646, A129}

\bibitem[\protect\citeauthoryear{{Johnson} et~al.,}{{Johnson}
  et~al.}{2014}]{2014MNRAS.444.3926J}
{Johnson} A.,  et~al., 2014, \mn@doi [\mnras] {10.1093/mnras/stu1615}, \href
  {https://ui.adsabs.harvard.edu/abs/2014MNRAS.444.3926J} {444, 3926}

\bibitem[\protect\citeauthoryear{{Koda} et~al.,}{{Koda}
  et~al.}{2014}]{2014MNRAS.445.4267K}
{Koda} J.,  et~al., 2014, \mn@doi [\mnras] {10.1093/mnras/stu1610}, \href
  {https://ui.adsabs.harvard.edu/abs/2014MNRAS.445.4267K} {445, 4267}

\bibitem[\protect\citeauthoryear{{Koda}, {Blake}, {Beutler}, {Kazin}  \&
  {Marin}}{{Koda} et~al.}{2016}]{2016MNRAS.459.2118K}
{Koda} J.,  {Blake} C.,  {Beutler} F.,  {Kazin} E.,   {Marin} F.,  2016,
  \mn@doi [\mnras] {10.1093/mnras/stw763}, \href
  {https://ui.adsabs.harvard.edu/abs/2016MNRAS.459.2118K} {459, 2118}

\bibitem[\protect\citeauthoryear{{Krause} \& {Eifler}}{{Krause} \&
  {Eifler}}{2017}]{2017MNRAS.470.2100K}
{Krause} E.,  {Eifler} T.,  2017, \mn@doi [\mnras] {10.1093/mnras/stx1261},
  \href {https://ui.adsabs.harvard.edu/abs/2017MNRAS.470.2100K} {470, 2100}

\bibitem[\protect\citeauthoryear{{Lai}, {Howlett}  \& {Davis}}{{Lai}
  et~al.}{2023}]{2023MNRAS.518.1840L}
{Lai} Y.,  {Howlett} C.,   {Davis} T.~M.,  2023, \mn@doi [\mnras]
  {10.1093/mnras/stac3252}, \href
  {https://ui.adsabs.harvard.edu/abs/2023MNRAS.518.1840L} {518, 1840}

\bibitem[\protect\citeauthoryear{{Lewis}, {Challinor}  \& {Lasenby}}{{Lewis}
  et~al.}{2000}]{2000ApJ...538..473L}
{Lewis} A.,  {Challinor} A.,   {Lasenby} A.,  2000, \mn@doi [\apj]
  {10.1086/309179}, \href
  {https://ui.adsabs.harvard.edu/abs/2000ApJ...538..473L} {538, 473}

\bibitem[\protect\citeauthoryear{{Li}, {Singh}, {Yu}, {Feng}  \& {Seljak}}{{Li}
  et~al.}{2019}]{2019JCAP...01..016L}
{Li} Y.,  {Singh} S.,  {Yu} B.,  {Feng} Y.,   {Seljak} U.,  2019, \mn@doi
  [\jcap] {10.1088/1475-7516/2019/01/016}, \href
  {https://ui.adsabs.harvard.edu/abs/2019JCAP...01..016L} {2019, 016}

\bibitem[\protect\citeauthoryear{{Lyall}, {Blake}, {Turner}, {Ruggeri}  \&
  {Winther}}{{Lyall} et~al.}{2023}]{2023MNRAS.518.5929L}
{Lyall} S.,  {Blake} C.,  {Turner} R.,  {Ruggeri} R.,   {Winther} H.,  2023,
  \mn@doi [\mnras] {10.1093/mnras/stac3323}, \href
  {https://ui.adsabs.harvard.edu/abs/2023MNRAS.518.5929L} {518, 5929}

\bibitem[\protect\citeauthoryear{{Mohammad} \& {Percival}}{{Mohammad} \&
  {Percival}}{2022}]{2022MNRAS.514.1289M}
{Mohammad} F.~G.,  {Percival} W.~J.,  2022, \mn@doi [\mnras]
  {10.1093/mnras/stac1458}, \href
  {https://ui.adsabs.harvard.edu/abs/2022MNRAS.514.1289M} {514, 1289}

\bibitem[\protect\citeauthoryear{{Norberg}, {Baugh}, {Gazta{\~n}aga}  \&
  {Croton}}{{Norberg} et~al.}{2009}]{2009MNRAS.396...19N}
{Norberg} P.,  {Baugh} C.~M.,  {Gazta{\~n}aga} E.,   {Croton} D.~J.,  2009,
  \mn@doi [\mnras] {10.1111/j.1365-2966.2009.14389.x}, \href
  {https://ui.adsabs.harvard.edu/abs/2009MNRAS.396...19N} {396, 19}

\bibitem[\protect\citeauthoryear{{Nusser}}{{Nusser}}{2017}]{2017MNRAS.470..445N}
{Nusser} A.,  2017, \mn@doi [\mnras] {10.1093/mnras/stx1225}, \href
  {https://ui.adsabs.harvard.edu/abs/2017MNRAS.470..445N} {470, 445}

\bibitem[\protect\citeauthoryear{{O'Connell} \& {Eisenstein}}{{O'Connell} \&
  {Eisenstein}}{2019}]{2019MNRAS.487.2701O}
{O'Connell} R.,  {Eisenstein} D.~J.,  2019, \mn@doi [\mnras]
  {10.1093/mnras/stz1359}, \href
  {https://ui.adsabs.harvard.edu/abs/2019MNRAS.487.2701O} {487, 2701}

\bibitem[\protect\citeauthoryear{{O'Connell}, {Eisenstein}, {Vargas}, {Ho}  \&
  {Padmanabhan}}{{O'Connell} et~al.}{2016}]{2016MNRAS.462.2681O}
{O'Connell} R.,  {Eisenstein} D.,  {Vargas} M.,  {Ho} S.,   {Padmanabhan} N.,
  2016, \mn@doi [\mnras] {10.1093/mnras/stw1821}, \href
  {https://ui.adsabs.harvard.edu/abs/2016MNRAS.462.2681O} {462, 2681}

\bibitem[\protect\citeauthoryear{{Okumura}, {Seljak}, {Vlah}  \&
  {Desjacques}}{{Okumura} et~al.}{2014}]{2014JCAP...05..003O}
{Okumura} T.,  {Seljak} U.,  {Vlah} Z.,   {Desjacques} V.,  2014, \mn@doi
  [\jcap] {10.1088/1475-7516/2014/05/003}, \href
  {https://ui.adsabs.harvard.edu/abs/2014JCAP...05..003O} {2014, 003}

\bibitem[\protect\citeauthoryear{{Park}}{{Park}}{2000}]{2000MNRAS.319..573P}
{Park} C.,  2000, \mn@doi [\mnras] {10.1046/j.1365-8711.2000.03886.x}, \href
  {https://ui.adsabs.harvard.edu/abs/2000MNRAS.319..573P} {319, 573}

\bibitem[\protect\citeauthoryear{{Philcox}, {Eisenstein}, {O'Connell}  \&
  {Wiegand}}{{Philcox} et~al.}{2020}]{2020MNRAS.491.3290P}
{Philcox} O. H.~E.,  {Eisenstein} D.~J.,  {O'Connell} R.,   {Wiegand} A.,
  2020, \mn@doi [\mnras] {10.1093/mnras/stz3218}, \href
  {https://ui.adsabs.harvard.edu/abs/2020MNRAS.491.3290P} {491, 3290}

\bibitem[\protect\citeauthoryear{{Said}, {Colless}, {Magoulas}, {Lucey}  \&
  {Hudson}}{{Said} et~al.}{2020}]{2020MNRAS.497.1275S}
{Said} K.,  {Colless} M.,  {Magoulas} C.,  {Lucey} J.~R.,   {Hudson} M.~J.,
  2020, \mn@doi [\mnras] {10.1093/mnras/staa2032}, \href
  {https://ui.adsabs.harvard.edu/abs/2020MNRAS.497.1275S} {497, 1275}

\bibitem[\protect\citeauthoryear{{Satpathy} et~al.,}{{Satpathy}
  et~al.}{2017}]{2017MNRAS.469.1369S}
{Satpathy} S.,  et~al., 2017, \mn@doi [\mnras] {10.1093/mnras/stx883}, \href
  {https://ui.adsabs.harvard.edu/abs/2017MNRAS.469.1369S} {469, 1369}

\bibitem[\protect\citeauthoryear{{Saulder}, {Mieske}, {Zeilinger}  \&
  {Chilingarian}}{{Saulder} et~al.}{2013}]{2013A&A...557A..21S}
{Saulder} C.,  {Mieske} S.,  {Zeilinger} W.~W.,   {Chilingarian} I.,  2013,
  \mn@doi [\aap] {10.1051/0004-6361/201321466}, \href
  {https://ui.adsabs.harvard.edu/abs/2013A&A...557A..21S} {557, A21}

\bibitem[\protect\citeauthoryear{{Saulder} et~al.,}{{Saulder}
  et~al.}{2023}]{2023arXiv230213760S}
{Saulder} C.,  et~al., 2023, \mn@doi [arXiv e-prints]
  {10.48550/arXiv.2302.13760}, \href
  {https://ui.adsabs.harvard.edu/abs/2023arXiv230213760S} {p. arXiv:2302.13760}

\bibitem[\protect\citeauthoryear{{Smith} et~al.,}{{Smith}
  et~al.}{2003}]{2003MNRAS.341.1311S}
{Smith} R.~E.,  et~al., 2003, \mn@doi [\mnras]
  {10.1046/j.1365-8711.2003.06503.x}, \href
  {https://ui.adsabs.harvard.edu/abs/2003MNRAS.341.1311S} {341, 1311}

\bibitem[\protect\citeauthoryear{{Springob} et~al.,}{{Springob}
  et~al.}{2014}]{2014MNRAS.445.2677S}
{Springob} C.~M.,  et~al., 2014, \mn@doi [\mnras] {10.1093/mnras/stu1743},
  \href {https://ui.adsabs.harvard.edu/abs/2014MNRAS.445.2677S} {445, 2677}

\bibitem[\protect\citeauthoryear{{Strauss} \& {Willick}}{{Strauss} \&
  {Willick}}{1995}]{1995PhR...261..271S}
{Strauss} M.~A.,  {Willick} J.~A.,  1995, \mn@doi [\physrep]
  {10.1016/0370-1573(95)00013-7}, \href
  {https://ui.adsabs.harvard.edu/abs/1995PhR...261..271S} {261, 271}

\bibitem[\protect\citeauthoryear{{Taylor} et~al.,}{{Taylor}
  et~al.}{2023}]{2023Msngr.190...46T}
{Taylor} E.~N.,  et~al., 2023, \mn@doi [The Messenger]
  {10.18727/0722-6691/5312}, \href
  {https://ui.adsabs.harvard.edu/abs/2023Msngr.190...46T} {190, 46}

\bibitem[\protect\citeauthoryear{{Tully}, {Courtois}  \& {Sorce}}{{Tully}
  et~al.}{2016}]{2016AJ....152...50T}
{Tully} R.~B.,  {Courtois} H.~M.,   {Sorce} J.~G.,  2016, \mn@doi [\aj]
  {10.3847/0004-6256/152/2/50}, \href
  {https://ui.adsabs.harvard.edu/abs/2016AJ....152...50T} {152, 50}

\bibitem[\protect\citeauthoryear{{Tully} et~al.,}{{Tully}
  et~al.}{2023}]{2023ApJ...944...94T}
{Tully} R.~B.,  et~al., 2023, \mn@doi [\apj] {10.3847/1538-4357/ac94d8}, \href
  {https://ui.adsabs.harvard.edu/abs/2023ApJ...944...94T} {944, 94}

\bibitem[\protect\citeauthoryear{{Turner}, {Blake}  \& {Ruggeri}}{{Turner}
  et~al.}{2021}]{2021MNRAS.502.2087T}
{Turner} R.~J.,  {Blake} C.,   {Ruggeri} R.,  2021, \mn@doi [\mnras]
  {10.1093/mnras/stab212}, \href
  {https://ui.adsabs.harvard.edu/abs/2021MNRAS.502.2087T} {502, 2087}

\bibitem[\protect\citeauthoryear{{Turner}, {Blake}  \& {Ruggeri}}{{Turner}
  et~al.}{2023}]{2023MNRAS.518.2436T}
{Turner} R.~J.,  {Blake} C.,   {Ruggeri} R.,  2023, \mn@doi [\mnras]
  {10.1093/mnras/stac3256}, \href
  {https://ui.adsabs.harvard.edu/abs/2023MNRAS.518.2436T} {518, 2436}

\bibitem[\protect\citeauthoryear{{Wadekar}, {Ivanov}  \&
  {Scoccimarro}}{{Wadekar} et~al.}{2020}]{2020PhRvD.102l3521W}
{Wadekar} D.,  {Ivanov} M.~M.,   {Scoccimarro} R.,  2020, \mn@doi [\prd]
  {10.1103/PhysRevD.102.123521}, \href
  {https://ui.adsabs.harvard.edu/abs/2020PhRvD.102l3521W} {102, 123521}

\makeatother
\end{thebibliography}

\appendix

\section{Derivation of velocity correlation tensor in terms of isotropic functions}
\label{sec:psiderivation}

In this Appendix we derive the expression for the velocity correlation tensor in terms of the isotropic functions $\psi_\parallel(r)$ and $\psi_\perp(r)$ defined by \cite{1988ApJ...332L...7G}.  We start from the definition of the correlation tensor (Eq.\ref{eq:velcorrtens}) in the form,
\begin{equation}
    \psi_{ij}(\vr) = \intk \, \frac{k_i \, k_j}{k^2} \, P_{vv}(\vk) \, e^{-i \vk \cdot \vr}
\end{equation}
Using the fact that $P_{vv}(\vk)$ is isotropic, and the plane wave expansion of $e^{-i \vk \cdot \vr}$, we find that
\begin{equation}
\begin{split}
    \psi_{ij}(\vr) &= \intkave \, P_{vv}(k) \sum_\ell (2\ell+1) \, i^\ell \, j_\ell(kr) \intok \, \hk_i \, \hk_j \, L_\ell(\hk \cdot \hr) \\
    &= \intkave \, P_{vv}(k) \sum_\ell (2\ell+1) \, i^\ell \, j_\ell(kr) \intok L_\ell(\hk \cdot \hr) \, L_1(\hk \cdot \hn_i) \, L_1(\hk \cdot \hn_j) 
\end{split}
\end{equation}
where $\hn_i$ is a unit vector in the $i$-direction.  Using the identity,
\begin{equation}
    \intok \, L_\ell(\hx \cdot \hk) \, L_1(\hy \cdot \hk) \, L_1(\hz \cdot \hk) = \frac{\delta^K_{l0}}{3} ( \hy \cdot \hz ) - \frac{\delta^K_{l2}}{15} \left[ ( \hy \cdot \hz ) - 3 ( \hx \cdot \hy ) ( \hx \cdot \hz ) \right] ,
\end{equation}
this relation becomes,
\begin{equation}
\begin{split}
    \psi_{ij}(\vr) &= \intkave \, P_{vv}(k) \sum_\ell (2\ell+1) \, i^\ell \, j_\ell(kr) \left\{ \frac{\delta^K_{l0}}{3} ( \hn_i \cdot \hn_j ) - \frac{\delta^K_{l2}}{15} \left[ ( \hn_i \cdot \hn_j ) - 3 ( \hr \cdot \hn_i ) ( \hr \cdot \hn_j ) \right] \right\} \\
    &= \intkave \, P_{vv}(k) \left\{ \frac{1}{3} \left[ j_0(kr) + j_2(kr) \right] \delta^K_{ij} - j_2(kr) \, \hr_i \, \hr_j \right\} \\
    &= \psi_\perp(r) \, \delta^K_{ij} + \left[ \psi_\parallel(r) - \psi_\perp(r) \right] \, \left( \frac{r_i \, r_j}{r^2} \right)
\end{split}
\end{equation}
using the Bessel function recursion relation, $j_0(x) + j_2(x) = \frac{3}{x} \, j_1(x)$.

\section{General expressions for covariance}
\label{sec:allthecovs}

\subsection{3D velocity statistics with vector separations}

In this section we provide the covariance relations between estimates of the ``simulation'' correlation functions $[ \hat{\xi}_{gg}, \hat{\xi}_{gv}, \hat{\xi}_{vv} ]$, which utilise the galaxy overdensity and inward velocity of one tracer towards the second tracer.  Here, we express these relations in a form preserving the direction of the two separations, $\vr$ and $\vs$, and we neglect redshift-space distortions. These covariance relations are given in terms of the dimensionless selection function of the density and velocity samples, $f_g(\vx) = w_g(\vx) \, n_g(\vx) \, V$ and $f_v(\vx) = w_v(\vx) \, n_v(\vx) \, V$, and their noise functions, $\sigma_g(\vx) = w_g(\vx) \sqrt{n_g(\vx) \, V}$ and $\sigma_v(\vx) = \epsilon_v(\vx) \, w_v(\vx) \sqrt{n_v(\vx) \, V}$, which are written in terms of the sample number densities $n_g(\vx)$ and $n_v(\vx)$, weight functions $w_g(\vx)$ and $w_v(\vx)$, individual-tracer velocity error $\epsilon_v(\vx)$, and Fourier cuboid volume $V$.  In this case, the velocity errors are applied independently to each velocity component.  These relations also use the galaxy power spectrum $P_{gg}(k) = b^2 \, P_m(k)$, the velocity power spectrum $P_{vv}(k) = \frac{a^2 H^2 f^2}{k^2} \, P_m(k)$, and the galaxy-velocity cross-power spectrum, $P_{gv}(k) = \frac{b a H f}{k} \, P_m(k)$.  To preserve the dimensionless nature of the integrand over $d^3\vx$, we assume that all these power spectra have been divided by $V$ before being used in these equations.  The normalisation constants are $N_{\xi_{gg}} = \intx f^2_g(\vx)$, $N_{\xi_{gv}} = \intx f_g(\vx) \, f_v(\vx)$ and $N_{\xi_{vv}} = \intx f^2_v(\vx)$.  The covariance relations are:
\begin{equation}
    {\rm Cov} \left[ \hat{\xi}_{gg}(\vr), \hat{\xi}_{gg}(\vs) \right] = N_{\xi_{gg}}^2 \intkv e^{-i \vk \cdot (\vr - \vs)} \intx \, \left[ f^2_g(\vx) \, P_{gg}(\vk) + \sigma_g^2(\vx) \right]^2
\end{equation}
\begin{equation}
    {\rm Cov} \left[ \hat{\xi}_{vv}(\vr), \hat{\xi}_{vv}(\vs) \right] = N_{\xi_{vv}}^2 \intkv e^{-i \vk \cdot (\vr - \vs)} \intx \, \left[ f^2_v(\vx) \, P_{vv}(\vk) \, ( \hr \cdot \hk ) \, ( \hs \cdot \hk ) + \sigma_v^2(\vx) \, ( \hr \cdot \hs ) \right]^2
\end{equation}
\begin{equation}
\begin{split}
    {\rm Cov} \left[ \hat{\xi}_{gv}(\vr), \hat{\xi}_{gv}(\vs) \right] = N_{\xi_{gv}}^2 & \intkv e^{-i \vk \cdot (\vr - \vs)} \intx \frac{1}{2} \left\{ \left[ f^2_g(\vx) \, P_{gg}(\vk) + \sigma_g^2(\vx) \right] \left[ f^2_v(\vx) \, P_{vv}(\vk) \, ( \hr \cdot \hk ) \, ( \hs \cdot \hk ) + \sigma_v^2(\vx) \, ( \hr \cdot \hs ) \right] \right. \\ &+ \left. f^2_g(\vx) \, f^2_v(\vx) \, P_{gv}^2(\vk) \, ( \hr \cdot \hk ) \, ( \hs \cdot \hk ) \right\}
\end{split}
\end{equation}
\begin{equation}
    {\rm Cov} \left[ \hat{\xi}_{gg}(\vr), \hat{\xi}_{vv}(\vs) \right] = N_{\xi_{gg}} N_{\xi_{vv}} \intkv e^{-i \vk \cdot (\vr - \vs)} \intx \, f^2_g(\vx) \, f^2_v(\vx) \, P_{gv}^2(\vk) \, ( \hs \cdot \hk )^2
\end{equation}
\begin{equation}
    {\rm Cov} \left[ \hat{\xi}_{gg}(\vr), \hat{\xi}_{gv}(\vs) \right] = N_{\xi_{gg}} N_{\xi_{gv}} \intkv e^{-i \vk \cdot (\vr - \vs)} \intx \, \left[ f^2_g(\vx) \, P_{gg}(\vk) + \sigma_g^2(\vx) \right] \, f_g(\vx) \, f_v(\vx) \, P_{gv}(\vk) \, ( \hr \cdot \hk )
\end{equation}
\begin{equation}
    {\rm Cov} \left[ \hat{\xi}_{gv}(\vr), \hat{\xi}_{vv}(\vs) \right] = N_{\xi_{gv}} N_{\xi_{vv}} \intkv e^{-i \vk \cdot (\vr - \vs)} \intx \left[ f^2_v(\vx) \, P_{vv}(\vk) \, ( \hr \cdot \hk ) \, ( \hs \cdot \hk ) + \sigma_v^2(\vx) \, ( \hr \cdot \hs ) \right] f_g(\vx) \, f_v(\vx) \, P_{gv}(\vk) \, ( \hs \cdot \hk )
\end{equation}

\subsection{Line-of-sight velocity statistics with vector separations}

In this section we provide the covariance relations between estimates of the ``observational'' correlation functions $[ \hat{\xi}_{gg} , \hat{\psi}_1 , \hat{\psi}_2 , \hat{\psi}_3 ]$, which utilise the galaxy overdensity and line-of-sight galaxy velocities.  Here, we express these relations in a form preserving the direction of the two separations, $\vr$ and $\vs$, and we neglect redshift-space distortions.  These relations use the same notation as the preceding section except in this case, the velocity errors are applied independently to each velocity component.  The normalisation constants are $N_{\psi_1} = \frac{1}{\langle \cos^2{\theta_{12}} \rangle}$, $N_{\psi_2} = \frac{1}{\langle \cos{\theta_{12}} \, \cos{\theta_1} \, \cos{\theta_2} \rangle}$ and $N_{\psi_3} = \frac{1}{\langle \cos^2{\theta_{12}} \rangle}$, where these are volume averages over the separation angles between the two line-of-sight directions, and between these two directions and the separation vector, as discussed in Sec.\ref{sec:lineofsight}.  The covariance relations are:
\begin{equation}
    {\rm Cov} \left[ \hat{\psi}_1(\vr), \hat{\psi}_1(\vs) \right] = N_{\psi_1}^2 \intkv e^{-i \vk \cdot (\vr - \vs)} \intx \, \left[ f^2_v(\vx) \, P_{vv}(\vk) \, ( \hx \cdot \hk )^2 + \sigma_v^2(\vx) \right]^2
\end{equation}
\begin{equation}
    {\rm Cov} \left[ \hat{\psi}_2(\vr), \hat{\psi}_2(\vs) \right] = N_{\psi_2}^2 \intkv e^{-i \vk \cdot (\vr - \vs)} \intx \, (\hx \cdot \hr)^2 \, (\hx \cdot \hs)^2 \left[ f^2_v(\vx) \, P_{vv}(\vk) \, ( \hx \cdot \hk )^2 + \sigma_v^2(\vx) \right]^2
\end{equation}
\begin{equation}
\begin{split}
    {\rm Cov} \left[ \hat{\psi}_3(\vr), \hat{\psi}_3(\vs) \right] = N_{\psi_3}^2 & \intkv e^{-i \vk \cdot (\vr - \vs)} \intx \frac{1}{2} \, (\hx \cdot \hr) \, (\hx \cdot \hs) \left\{ \left[ f^2_g(\vx) \, P_{gg}(\vk) + \sigma_g^2(\vx) \right] \left[ f^2_v(\vx) \, P_{vv}(\vk) \, ( \hx \cdot \hk )^2 + \sigma_v^2(\vx) \right] \right. \\ &+ \left. f^2_g(\vx) \, f^2_v(\vx) \, P_{gv}^2(\vk) \, ( \hx \cdot \hk )^2 \right\}
\end{split}
\end{equation}
\begin{equation}
    {\rm Cov} \left[ \hat{\psi}_1(\vr), \hat{\psi}_2(\vs) \right] = N_{\psi_1} N_{\psi_2} \intkv e^{-i \vk \cdot (\vr - \vs)} \intx \, (\hx \cdot \hs)^2 \left[ f^2_v(\vx) \, P_{vv}(\vk) \, ( \hx \cdot \hk )^2 + \sigma_v^2(\vx) \right]^2
\end{equation}
\begin{equation}
    {\rm Cov} \left[ \hat{\psi}_1(\vr), \hat{\psi}_3(\vs) \right] = N_{\psi_1} N_{\psi_3} \intkv e^{-i \vk \cdot (\vr - \vs)} \intx \, \left[ f^2_v(\vx) \, P_{vv}(\vk) \, ( \hx \cdot \hk )^2 + \sigma_v^2(\vx) \right] \, f_g(\vx) \, f_v(\vx) \, P_{gv}(\vk) \, ( \hx \cdot \hk )
\end{equation}
\begin{equation}
    {\rm Cov} \left[ \hat{\psi}_2(\vr), \hat{\psi}_3(\vs) \right] = N_{\psi_2} N_{\psi_3} \intkv e^{-i \vk \cdot (\vr - \vs)} \intx (\hx \cdot \hr)^2 (\hx \cdot \hs)^2 \left[ f^2_v(\vx) P_{vv}(\vk) ( \hx \cdot \hk )^2 + \sigma_v^2(\vx) \right] f_g(\vx) f_v(\vx) P_{gv}(\vk) ( \hx \cdot \hk )
\end{equation}
\begin{equation}
    {\rm Cov} \left[ \hat{\xi}_{gg}(\vr), \hat{\psi}_1(\vs) \right] = N_{\xi_{gg}} N_{\psi_1} \intkv e^{-i \vk \cdot (\vr - \vs)} \intx \, f_g(\vx)^2 \, f_v(\vx)^2 \, P_{gv}(\vk)^2 \, ( \hx \cdot \hk )^2
\end{equation}
\begin{equation}
    {\rm Cov} \left[ \hat{\xi}_{gg}(\vr), \hat{\psi}_2(\vs) \right] = N_{\xi_{gg}} N_{\psi_2} \intkv e^{-i \vk \cdot (\vr - \vs)} \intx \, f_g(\vx)^2 \, f_v(\vx)^2 \, P_{gv}(\vk)^2 \, ( \hx \cdot \hk )^2 \, ( \hx \cdot \hs )^2
\end{equation}
\begin{equation}
    {\rm Cov} \left[ \hat{\xi}_{gg}(\vr), \hat{\psi}_3(\vs) \right] = N_{\xi_{gg}} N_{\psi_3} \intkv e^{-i \vk \cdot (\vr - \vs)} \intx \, \left[ f^2_g(\vx) \, P_{gg}(\vk) + \sigma_g^2(\vx) \right] \, f_g(\vx) \, f_v(\vx) \, P_{gv}(\vk) \, ( \hx \cdot \hk ) \, ( \hx \cdot \hs )
\end{equation}

\subsection{Angle-averaged 3D velocity statistics}

In this section we provide the covariance relations when the estimators of the ``simulation'' correlation functions $[ \hat{\xi}_{gg} , \hat{\psi}_1 , \hat{\psi}_2 , \hat{\psi}_3 ]$, are averaged over the direction of the separations.  When taking these angle averages, we introduce the volume-averaged quantities $W_n(f) = \intx \, f(\vx) \intok ( \hx \cdot \hk )^n$, where $f$ represents a combination of the selection function or noise fields.  The power spectra $(P_{gg}, P_{gv}, P_{vv})$ are all functions of $k$, and are assumed divided by the Fourier cuboid volume $V$.  The covariance relations are:
\begin{equation}
    {\rm Cov} \left[ \hat{\xi}_{gg}(r) , \hat{\xi}_{gg}(s) \right] = N_{\xi_{gg}}^2 \intkvave \, j_0(kr) \, j_0(ks) \left[ W_0(f_g^4) \, P_{gg}^2 + 2 \, W_0(f_g^2 \sigma_g^2) \, P_{gg} + W_0(\sigma_g^4) \right]
\end{equation}
\begin{equation}
\begin{split}
    {\rm Cov} \left[ \hat{\xi}_{vv}(r) , \hat{\xi}_{vv}(s) \right] = N_{\xi_{vv}}^2 & \intkvave \left\{ \left[ j_0(kr) - \frac{2 j_1(kr)}{kr} \right] \left[ j_0(ks) - \frac{2 j_1(ks)}{ks} \right] \left[ W_0(f_v^4) \, P_{vv}^2 + 2 \, W_0(f_v^2 \sigma_v^2) \, P_{vv} + W_0(\sigma_v^4) \right] \right. \\ &+ \left. 2 \, \frac{j_1(kr)}{kr} \frac{j_1(ks)}{ks} \, W_0(\sigma_v^4) \right\}
\end{split}
\end{equation}
\begin{equation}
\begin{split}
    {\rm Cov} \left[ \hat{\xi}_{gv}(r) , \hat{\xi}_{gv}(s) \right] = N_{\xi_{gv}}^2 & \intkvave \, j_1(kr) \, j_1(ks) \\ & \frac{1}{2} \left[ W_0(f_g^2 f_v^2) \left( P_{gg} \, P_{vv} + P_{gv}^2 \right) + W_0(f_v^2 \sigma_g^2) \, P_{vv} + W_0(f_g^2 \sigma_v^2) \, P_{gg} + W_0(\sigma_g^2 \sigma_v^2) \right]
\end{split}
\end{equation}
\begin{equation}
    {\rm Cov} \left[ \hat{\xi}_{gg}(r) , \hat{\xi}_{vv}(s) \right] = N_{\xi_{gg}} N_{\xi_{vv}} \intkvave \, j_0(kr) \, \left[ j_0(ks) - \frac{2 j_1(ks)}{ks} \right] \, W_0(f_g^2 f_v^2) \, P_{gv}^2
\end{equation}
\begin{equation}
    {\rm Cov} \left[ \hat{\xi}_{gg}(r) , \hat{\xi}_{gv}(s) \right] = N_{\xi_{gg}} \, N_{\xi_{gv}} \intkvave \, j_0(kr) \, j_1(ks) \, \left[ W_0(f_g^3 f_v) \, P_{gg} \, P_{gv} + W_0(f_g f_v \sigma_g^2) \, P_{gv} \right]
\end{equation}
\begin{equation}
    {\rm Cov} \left[ \hat{\xi}_{gv}(r) , \hat{\xi}_{vv}(s) \right] = N_{\xi_{gv}} \, N_{\xi_{vv}} \, \intkvave \, j_1(kr) \, \left[ j_0(ks) - \frac{2 j_1(ks)}{ks} \right] \, \left[ W_0(f_g f_v^3) \, P_{vv} \, P_{gv} + W_0(f_g f_v \sigma_v^2) \, P_{gv} \right]
\end{equation}

\subsection{Angle-averaged line-of-sight velocity statistics}

In this section we provide the covariance relations when the estimators of the ``observational'' correlation functions $[ \hat{\xi}_{gg} , \hat{\psi}_1 , \hat{\psi}_2 , \hat{\psi}_3 ]$ are averaged over the direction of the separations.  We neglect redshift-space distortions and use the same notation as the preceding section.  The covariance relations are:
\begin{equation}
    {\rm Cov} \left[ \hat{\psi}_1(r) , \hat{\psi}_1(s) \right] = N_{\psi_1}^2 \intkvave \, j_0(kr) \, j_0(ks) \left[ W_4(f_v^4) \, P_{vv}^2 + 2 W_2(f_v^2 \sigma_v^2) \, P_{vv} + W_0(\sigma_v^4) \right]
\end{equation}
\begin{equation}
\begin{split}
    {\rm Cov} \left[ \hat{\psi}_2(r) , \hat{\psi}_2(s) \right] = N_{\psi_2}^2 & \intkvave \left\{ \frac{j_1(kr)}{kr} \, \frac{j_1(ks)}{ks} \left[ W_4(f_v^4) \, P_{vv}^2 + 2 W_2(f_v^2 \sigma_v^2) \, P_{vv} + W_0(\sigma_v^4) \right] \right. \\
    &+ \left( \frac{j_1(kr)}{kr} j_2(ks) + \frac{j_1(ks)}{ks} j_2(kr) \right) \left[ W_6(f_v^4) \, P_{vv}^2 + 2 W_4(f_v^2 \sigma_v^2) \, P_{vv} + W_2(\sigma_v^4) \right] \\
    &+ \left. j_2(kr) j_2(ks) \left[ W_8(f_v^4) \, P_{vv}^2 + 2 W_6(f_v^2 \sigma_v^2) \, P_{vv} + W_4(\sigma_v^4) \right] \right\}
\end{split}
\end{equation}
\begin{equation}
\begin{split}
    {\rm Cov} \left[ \hat{\psi}_3(r) , \hat{\psi}_3(s) \right] = N_{\psi_3}^2 & \intkvave \, j_1(kr) \, j_1(ks) \frac{1}{2} \left\{ W_4(f_g^2 f_v^2) \left( P_{gg} \, P_{vv} + P_{gv}^2 \right) + W_4(f_v^2 \sigma_g^2) \, P_{vv} \right. \\
    &+ \left. W_2(f_g^2 \sigma_v^2) \, P_{gg} + W_2(\sigma_g^2 \sigma_v^2) \right\}
\end{split}
\end{equation}
\begin{equation}
\begin{split}
    {\rm Cov} \left[ \hat{\psi}_1(r) , \hat{\psi}_2(s) \right] = N_{\psi_1} N_{\psi_2} & \intkvave \, \left\{ j_0(kr) \, \frac{j_1(ks)}{ks} \left[ W_4(f_v^4) \, P_{vv}^2 + 2 W_2(f_v^2 \sigma_v^2) \, P_{vv} + W_0(\sigma_v^4) \right] \right. \\
    &+ \left. j_0(kr) \, j_2(ks) \left[ W_6(f_v^4) \, P_{vv}^2 + 2 W_4(f_v^2 \sigma_v^2) \, P_{vv} + W_2(\sigma_v^4) \right] \right\}
\end{split}
\end{equation}
\begin{equation}
    {\rm Cov} \left[ \hat{\psi}_1(r) , \hat{\psi}_3(s) \right] = N_{\psi_1} N_{\psi_3} \intkvave \, j_0(kr) \, j_1(ks) \left[ W_4(f_g f_v^3) P_{vv} P_{gv} + W_2(f_g f_v \sigma_v^2) P_{gv} \right]
\end{equation}
\begin{equation}
\begin{split}
    {\rm Cov} \left[ \hat{\psi}_2(r) , \hat{\psi}_3(s) \right] = N_{\psi_2} N_{\psi_3} & \intkvave \, \left\{ \frac{j_1(kr)}{kr} \, j_1(ks) \left[ W_4(f_g f_v^3) \, P_{vv} \, P_{gv} + W_2(f_g f_v \sigma_v^2) \, P_{gv} \right] \right. \\
    &- \left. j_2(kr) \, j_1(ks) \left[ W_6(f_g f_v^3) \, P_{vv} \, P_{gv} + W_4(f_g f_v \sigma_v^2) \, P_{gv} \right] \right\}
\end{split}
\end{equation}
\begin{equation}
    {\rm Cov} \left[ \hat{\xi}_{gg}(r) , \hat{\psi}_1(s) \right] = N_{\xi_{gg}} N_{\psi_1} \intkvave \, j_0(kr) \, j_0(ks) \, W_2(f_g^2 f_v^2) \, P^2_{gv}
\end{equation}
\begin{equation}
    {\rm Cov} \left[ \hat{\xi}_{gg}(r) , \hat{\psi}_2(s) \right] = N_{\xi_{gg}} N_{\psi_2} \intkvave \, \left\{ j_0(kr) \, \frac{j_1(ks)}{ks} \, W_2(f_g^2 f_v^2) \, P^2_{gv} - j_0(kr) \, j_2(ks) \, W_4(f_g^2 f_v^2) \, P^2_{gv} \right\} \, 
\end{equation}
\begin{equation}
    {\rm Cov} \left[ \hat{\xi}_{gg}(r) , \hat{\psi}_3(s) \right] = N_{\xi_{gg}} N_{\psi_3} \intkvave \, j_0(kr) \, j_1(ks) \, \left[ W_2(f_g^2 f_v^2) \, P_{gg} \, P_{gv} + W_2(f_g f_v \sigma_v^2) \, P_{gv} \right]
\end{equation}

\subsection{Covariances of multipoles}

In this section we provide the covariance relations when the estimators of the ``observational'' correlation functions including redshift-space distortions, $[ \hat{\xi}^0_{gg} , \hat{\xi}^2_{gg} , \hat{\xi}^1_{gu} , \hat{\psi}_1 , \hat{\psi}_2 ]$, are averaged over the direction of the separation.  When taking these angle averages, we introduce the volume-averaged quantities $V_n(f) = \intx \, f(\vx) \intok \left[ L_2( \hx \cdot \hk ) \right]^n$, where $f$ represents a combination of the selection function or noise fields.  In Eq.\ref{eq:covxi0ggpsi1} to Eq.\ref{eq:covxi2ggpsi2}, all averages $V_n(f)$ are taken over the field $f(\vx) = f^2_g(\vx) \, f^2_v(\vx)$, and this is not repeated inside the equations.  We also use the first two multipoles of the galaxy auto-power spectrum, $P^0_{gg}$ and $P^2_{gg}$, and the galaxy-velocity cross-power spectrum, $P^0_{gv}$ and $P^2_{gv}$, whose forms in linear theory are given in Eq.\ref{eq:pkpole}.  The subsequent relations are derived by substituting position-dependent power spectra for $P_{gg}(\vk)$ and $P_{gv}(\vk)$ in the preceding equations, for example,
\begin{equation}
    P_{gg}(\vk) \rightarrow P_{gg}(\vk,\vx) = \sum_\ell P_{gg}^\ell(k) \, L_\ell(\hk \cdot \hx) = P^0_{gg}(k) + L_2(\hk \cdot \hx) \, P^2_{gg}(k) .
\end{equation}
The following covariances result after averaging over angles:
\begin{equation}
    {\rm Cov} \left[ \hat{\xi}^0_{gg}(r) , \hat{\xi}^0_{gg}(s) \right] = N_{\xi_{gg}}^2 \intkvave \, j_0(kr) \, j_0(ks) \left\{ V_0(f_g^4) \, (P^0_{gg})^2 + V_2(f_g^4) \, (P^2_{gg})^2 + 2 \, V_0(f_g^2 \sigma_g^2) \, P^0_{gg} + V_0(\sigma_g^4) \right\}
\end{equation}
\begin{equation}
\begin{split}
    {\rm Cov} \left[ \hat{\xi}^2_{gg}(r) , \hat{\xi}^2_{gg}(s) \right] = 25 \, N_{\xi_{gg}}^2  & \intkvave j_2(kr) \, j_2(ks) \, \left\{ V_2(f_g^4) \, (P^0_{gg})^2 + 2 \, V_3(f_g^4) \, P^0_{gg} \, P^2_{gg} + V_4(f_g^4) \, (P^2_{gg})^2 \right. \\ &+ \left. 2 \, V_2(f_g^2 \sigma_g^2) \, P^0_{gg} + 2 \, V_3(f_g^2 \sigma_g^2) \, P^2_{gg} + V_2(\sigma_g^4) \right\}
\end{split}
\end{equation}
\begin{equation}
    {\rm Cov} \left[ \hat{\xi}^0_{gg}(r) , \hat{\xi}^2_{gg}(s) \right] = -5 \, N_{\xi_{gg}}^2 \intkvave j_0(kr) \, j_2(ks) \left\{ 2 \, V_2(f_g^4) \,  P^0_{gg} \, P^2_{gg} + V_3(f_g^4) \, (P^2_{gg})^2 + 2 \, V_2(f_g^2 \sigma_g^2) \, P^2_{gg} \right\}
\end{equation}.
\begin{equation}
    {\rm Cov} \left[ \hat{\xi}^0_{gg}(r) , \hat{\psi}_1(s) \right] = N_{\xi_{gg}} N_{\psi_1} \intkvave j_0(kr) \, j_0(ks) \left\{ \tfrac{1}{3} V_0 \, ( P^0_{gv} )^2 + \tfrac{4}{3} V_2 \, P^0_{gv} \, P^2_{gv} + \left( \tfrac{1}{3} V_2 + \tfrac{2}{3} V_3 \right) ( P^2_{gv} )^2 \right\}
\label{eq:covxi0ggpsi1}
\end{equation}
\begin{equation}
    {\rm Cov} \left[ \hat{\xi}^2_{gg}(r) , \hat{\psi}_1(s) \right] = -5 \, N_{\xi_{gg}} N_{\psi_1} \intkvave j_2(kr) \, j_0(ks) \left\{ \tfrac{2}{3} V_2 \, ( P^0_{gv} )^2 + \left( \tfrac{2}{3} V_2 + \tfrac{4}{3} V_3 \right) P^0_{gv} \, P^2_{gv} + \left( \tfrac{1}{3} V_3 + \tfrac{2}{3} V_4 \right) ( P^2_{gv} )^2 \right\}
\end{equation}
\begin{equation}
\begin{split}
    {\rm Cov} \left[ \hat{\xi}^0_{gg}(r) , \hat{\psi}_2(s) \right] = N_{\xi_{gg}} N_{\psi_2} & \intkvave \left\{ j_0(kr) \, \frac{j_1(ks)}{ks} \left[ \tfrac{1}{3} V_0 \, ( P^0_{gv} )^2 + \tfrac{4}{3} V_2 \, P^0_{gv} \, P^2_{gv} + \left( \tfrac{1}{3} V_2 + \tfrac{2}{3} V_3 \right) ( P^2_{gv} )^2 \right] \right. \\
    &\left. - j_0(kr) \, j_2(ks) \left[ \left( \tfrac{1}{9} V_0 + \tfrac{4}{9} V_2 \right) (P^0_{gv})^2 + \left( \tfrac{8}{9} V_2 + \tfrac{8}{9} V_3 \right) \, P^0_{gv} \, P^2_{gv} + \left( \tfrac{1}{9} V_2+ \tfrac{4}{9} V_3 + \tfrac{4}{9} V_4 \right) (P^2_{gv})^2 \right] \right\}
\end{split}
\end{equation}
\begin{equation}
\begin{split}
    {\rm Cov} \left[ \hat{\xi}^2_{gg}(r) , \hat{\psi}_2(s) \right] = -5 \, N_{\xi_{gg}} N_{\psi_2} & \intkvave \left\{ j_2(kr) \, \frac{j_1(ks)}{ks} \left[ \tfrac{2}{3} V_2 ( P^0_{gv} )^2 + \left( \tfrac{2}{3} V_2 + \tfrac{4}{3} V_3 \right) P^0_{gv} \, P^2_{gv} + \left( \tfrac{1}{3} V_3 + \tfrac{2}{3} V_4 \right) ( P^2_{gv} )^2 \right] \right. \\
    & - \left. j_2(kr) \, j_2(ks) \left[ \left( \tfrac{4}{9} V_2 + \tfrac{4}{9} V_3 \right) (P^0_{gv})^2 + \left( \tfrac{2}{9} V_2 + \tfrac{8}{9} V_3 + \tfrac{8}{9} V_4 \right) P^0_{gv} \, P^2_{gv} \right] + \left( \tfrac{1}{9} V_3 + \tfrac{4}{9} V_4 \right) (P^2_{gv})^2 \right\}
\end{split}
\label{eq:covxi2ggpsi2}
\end{equation}
\begin{equation}
\begin{split}
    {\rm Cov} \left[ \hat{\xi}^0_{gg}(r) , \hat{\xi}^1_{gu}(s) \right] = N_{\xi_{gg}} N_{\psi_3} & \intkvave j_0(kr) \, j_1(ks) \left\{ P^0_{gv} \left( \tfrac{1}{3} V_0(f_g^3 f_v) \, P^0_{gg} + \tfrac{2}{3} V_2(f_g^3 f_v) \, P^2_{gg} + \tfrac{1}{3} V_0(f_g f_v \sigma_g^2) \right) \right. \\
    &+ \left. P^2_{gv} \left( \tfrac{2}{3} V_2(f_g^3 f_v) \, P^0_{gg} + \tfrac{2}{3} V_2(f_g f_v \sigma_g^2) \right) + \left( \tfrac{1}{3} V_2(f_g^3 f_v) + \tfrac{2}{3} V_3(f_g^3 f_v) \right) P^2_{gg} \, P^2_{gv} \right\}
\end{split}
\end{equation}
\begin{equation}
\begin{split}
    {\rm Cov} \left[ \hat{\xi}^2_{gg}(r) , \hat{\xi}^1_{gu}(s) \right] = - 5 \, N_{\xi_{gg}} N_{\psi_3} & \intkvave j_2(kr) \, j_1(ks) \left\{ V_2(f_g^3 f_v) \left( \tfrac{2}{3} P^0_{gg} P^0_{gv} + \tfrac{1}{3} P^0_{gg} P^2_{gv} + \tfrac{1}{3} P^2_{gg} P^0_{gv} \right) \right. \\
    &+ V_3(f_g^3 f_v) \left( \tfrac{2}{3} P^0_{gg} P^2_{gv} + \tfrac{2}{3} P^2_{gg} P^0_{gv} + \tfrac{1}{3} P^2_{gg} P^2_{gv} \right) + \tfrac{2}{3} V_4(f_g^3 f_v) \, P^2_{gg} \, P^2_{gv} \\
    &+ \left. V_2(f_g f_v \sigma_v^2) \left( \tfrac{2}{3} P^0_{gv} + \tfrac{1}{3} P^2_{gv} \right) + \tfrac{2}{3} V_3(f_g f_v \sigma_v^2) \, P^2_{gv} \right\}
\end{split}
\end{equation}
\begin{equation}
\begin{split}
    {\rm Cov} \left[ \hat{\xi}^1_{gu}(r) , \hat{\xi}^1_{gu}(s) \right] = N_{\psi_3}^2 & \intkvave j_1(kr) \, j_1(ks) \, \frac{1}{2} \left\{ V_0(f_g^2 f_v^2) \left( \tfrac{1}{9} P^0_{gg} P^0_{vv} + \tfrac{1}{9} (P^0_{gv})^2 \right) \right. \\
    &+ V_2(f_g^2 f_v^2) \left( \tfrac{4}{9} P^0_{gg} P^0_{vv} + \tfrac{4}{9} (P^0_{gv})^2 + \tfrac{4}{9} P^0_{vv} P^2_{gg} + \tfrac{8}{9} P^0_{gv} P^2_{gv} + \tfrac{1}{9} (P^2_{gv})^2 \right) \\
    &+ V_3(f_g^2 f_v^2) \left( \tfrac{4}{9} P^0_{vv} P^2_{gg} + \tfrac{8}{9} P^0_{gv} P^2_{gv} + \tfrac{4}{9} (P^2_{gv})^2 \right) + \tfrac{4}{9} V_4(f_g^2 f_v^2) \, (P^2_{gv})^2 \\
    &+ \left. \left( \tfrac{1}{9} V_0(f_v^2 \sigma_g^2) + \tfrac{4}{9} V_2(f_v^2 \sigma_g^2) \right) \, P^0_{vv} + \tfrac{1}{3} V_0(f_g^2 \sigma_v^2) \, P^0_{gg} + \tfrac{2}{3} V_2(f_g^2 \sigma_v^2) \, P^2_{gg} + \tfrac{1}{3} V_0(\sigma_g^2 \sigma_v^2) \right\}
\end{split}
\end{equation}
\begin{equation}
\begin{split}
    {\rm Cov} \left[ \hat{\xi}^1_{gu}(r) , \hat{\psi}_1(s) \right] = N_{\psi_3} N_{\psi_1} & \intkvave j_1(kr) \, j_0(ks) \left\{ \left( \tfrac{1}{9} V_0(f_g f_v^3) + \tfrac{4}{9} V_2(f_g f_v^3) \right) P_{vv} \, P^0_{gv} + \tfrac{1}{3} V_0(f_g f_v \sigma_v^2) P^0_{gv} \right. \\
    & \left. \left( \tfrac{4}{9} V_2(f_g f_v^3) + \tfrac{4}{9} V_3(f_g f_v^3) \right) P_{vv} \, P^2_{gv} + \tfrac{2}{3} V_2(f_g f_v \sigma_v^2) \, P^2_{gv} \right\}
\end{split}
\end{equation}
\begin{equation}
\begin{split}
    {\rm Cov} \left[ \hat{\xi}^1_{gu}(r) , \hat{\psi}_2(s) \right] = N_{\psi_3} N_{\psi_2} & \intkvave \left\{ j_1(kr) \, \frac{j_1(ks)}{ks}  \left[ \left( \tfrac{1}{9} V_0(f_g f_v^3) + \tfrac{4}{9} V_2(f_g f_v^3) \right) P_{vv} \, P^0_{gv} + \tfrac{1}{3} V_0(f_g f_v \sigma_v^2) \, P^0_{gv} \right. \right. \\
    & \left. \left( \tfrac{4}{9} V_2(f_g f_v^3) + \tfrac{4}{9} V_3(f_g f_v^3) \right) P_{vv} \, P^2_{gv} + \tfrac{2}{3} V_2(f_g f_v \sigma_v^2) \, P^2_{gv} \right] \\
    & - j_1(kr) \, j_2(ks) \left[ \left( \tfrac{1}{27} V_0(f_g f_v^3) + \tfrac{4}{9} V_2(f_g f_v^3) + \tfrac{8}{27} V_3(f_g f_v^3) \right) P_{vv} \, P^0_{gv} \right. \\
    &+ \left( \tfrac{1}{9} V_0(f_g f_v \sigma_v^2) + \tfrac{4}{9} V_2(f_g f_v \sigma_v^2) \right) P^0_{gv} + \left( \tfrac{2}{9} V_2(f_g f_v^3) + \tfrac{4}{9} V_3(f_g f_v^3) + \tfrac{8}{27} V_4(f_g f_v^3) \right) P_{vv} \, P^2_{gv} \\
    & \left. \left. + \left( \tfrac{4}{9} V_2(f_g f_v \sigma_v^2) + \tfrac{4}{9} V_3(f_g f_v \sigma_v^2) \right) P^2_{gv} \right] \right\}
\end{split}
\end{equation}

\bsp
\label{lastpage}
\end{document}